# MINFF: A Family of Classical Force Fields for Minerals with Tunable Angle Terms

*Michael Holmboe*, Nathan Lavauzelle, Rogers Swai.*

Department of Chemistry, Umeå University, SE-901 87 Umeå, Sweden

ABSTRACT

Atomistic simulations of minerals are limited by the accuracy and transferability of the force field. We introduce MINFF, a force field family for layered and non-layered oxides and hydroxides of M being Si, Al, Fe, Mg, Ca, Ti, Li and F. MINFF has no M-O bond terms. Instead, M-O distances follow from electrostatics and dispersion, while a single harmonic angle term on all M-O triplets controls coordination geometry. Metal charges are DDEC6 charges from reference mineral structures, and oxygen charges are assigned by nearest-neighbour charge smearing, covering basal, apical, hydroxyl and edge sites and sites adjacent to isomorphous substitutions. The oxygen Lennard-Jones parameters are based on the OPC3 water model and change by less than 5% when re-optimised per mineral, so compatibility with OPC3 and related ion-pair potentials is measured rather than assumed. Metal Lennard-Jones parameters were fitted to unit-cell metrics, bond distances and angles, both synchronously across more than 30 reference minerals (general, GMINFF) and for individual minerals (tailored, TMINFF), each at four angle force constants (0, 250, 500 and 1500 kJ/mol/rad$^2$). With angle terms, MINFF reproduces experimental unit-cells metrics to within 0.5% and atom-type distances and angles to within 0.7-1.4% and 0.1-0.4%, with Global Instability Index values as low as 0.11, below those of CLAYFF (0.34) and INTERFACE (0.21). A moderate angle constant reproduces the stiff in-plane and soft shear elastic constants of phyllosilicates and improves the Kaolinite librational bands, and in hydrated Na-Montmorillonite MINFF gives the lowest residual factor against the 2W 00l pattern of the models tested. The general and tailored sets at four angle stiffnesses give direct control over the transferability-fidelity trade-off. Topologies can be generated with the atom (MATLAB) and atomipy (Python) libraries and at www.atomipy.io.

## 1. Introduction

Clay minerals and structurally related oxides and hydroxides, including phyllosilicates, simple metal oxides, layered double hydroxides, and microporous tectosilicates such as zeolites, are central to a remarkably broad set of geochemical, environmental, and industrial problems or applications. They control the transport and retention of water, ions, and contaminants in soils and sedimentary rocks, dominate the surface chemistry of weathering and biogeochemical cycling, and play a defining role in the long-term safety assessment of engineered barriers for radioactive waste disposal. The same families of minerals underpin major industrial processes, from cement and concrete chemistry, through catalysis and ion exchange in zeolites, to the design of clay nanocomposites, drilling fluids, drug-delivery vehicles, and emerging electrochemical interfaces. Their characteristic combination of layered or framework topology, large specific surface area, charge defects arising from isomorphic substitutions, and pH-dependent edge chemistry gives rise to many of these properties but also makes their atomic-scale behaviour difficult to access experimentally.

Classical molecular dynamics (MD) and Monte Carlo (MC) simulations have therefore become important complementary tools to X-ray diffraction, neutron scattering, NMR, infrared or synchrotron-based spectroscopies, providing molecular-scale insight into the structure, dynamics, and thermodynamics of mineral surfaces and interfaces, swelling clays, hydration behaviour, and confined fluids. The predictive power of such simulations, however, depends critically on the quality of the underlying force field, and its compatibility with the appropriate water model and ion-pair potentials and force fields of inorganic or organic adsorbates. For minerals, the dominant choices in the past two decades have been CLAYFF and the INTERFACE force fields,[1–4] together with a number of more specialised potentials for silica and zeolites (e.g. TraPPE-zeo).[5–8] MINFF was initially primarily developed for silicate minerals, for which both the CLAYFF and INTERFACE force fields have been used extensively and with considerable success. CLAYFF is a non-bonded force field with a single harmonic O–H term (and optional M–O–H angle terms), designed to be a simple and transferable force field for many minerals with a distinct set of atom-types, having oxygen Lennard-Jones parameters taken from and assuring the compatibility with the SPC-oxygen. Whereas the partial charges in CLAYFF are close to half-ionised values, INTERFACE instead presents substantially lower partial charges, necessitating explicit and firm bond and angle terms for silicate minerals, and targets especially organic–inorganic interactions compatible with CHARMM and PCFF systems and surface energies.

Due to their fundamental differences, especially in the magnitude of partial charges, CLAYFF and INTERFACE show systematic differences when benchmarked against many observables, for instance the symmetry of coordination geometries, and silicate sheet cavity relaxation, lattice strain, and vibrational properties, all of which propagate into derived quantities such as theoretical XRD profiles, basal plane density maps, immersion energies, and elastic constants.[9–14] Since CLAYFF leaves the M–O angles without any bonded interactions, it can produce qualitatively wrong coordination and sheet silicate geometries,[9,15] as well as unstable edge sites and for certain minerals yield Bond Valence Sum-derived Global Instability Indices ($G_{II}$) above the 0.2 threshold typically associated with chemically unstable lattices.

For INTERFACE, recent oxide/hydroxide force field development partially replaces some of the prior models,[3,16] but remains restricted to specific mineral models and does not provide a unified treatment of substitutions, undercoordinated edge groups, and zeolite frameworks within a single, mutually consistent parameter set.

Both CLAYFF and INTERFACE force fields have evolved through a sequence of modifications that, taken together, document the same gaps that motivate MINFF: an improved ability to handle undercoordinated edge sites, the maintenance of asymmetric and ditrigonal cavities, and the need to cover mineral systems beyond the original training sets.

On the CLAYFF side, the general modifications or new developments fall into four overlapping families: (i) tuning of basal-oxygen, or the cation-basal-oxygen non-bonded parameters, in order to improve the hydration properties and structure for substituted smectite clays,[17–19] (ii) addition of harmonic M–O–H angle-bending terms to stabilise edge and surface hydroxyls,[10–12,20] (iii) extensions to new mineral families and to high-pressure / high-temperature regimes,[21] and (iv) DDEC-charge silica derivatives with refit cross-LJ parameters for silica–water interfaces.[22,23] The umbrella reference for this trajectory is the CLAYFF review by Cygan, Greathouse & Kalinichev.[24]

On the INTERFACE side, an equally active programme has produced compound-specific extensions across silicates, silica, alumina, oxides and hydroxides, metals, sulfates, apatites, and, most recently, a reactive parameter set. Key milestones include the extended-Born charge methodology,[25] the original phyllosilicate parameterisation,[2] the umbrella feature article on thermodynamically consistent INTERFACE,[4] the silica force field with its pH-resolved surface-model database,[26] the broad oxide/hydroxide extension,[3] the most recent alumina and electrolyte-interface parameter set,[16] and the reactive INTERFACE force field INTERFACE-R that replaces harmonic bonds with Morse potentials.[27]

The MINFF family presented here is an attempt to address some of the recurring gaps of the above-mentioned force fields into a single, internally consistent framework: a general parameter set (GMINFF) optimised synchronously across more than 30 minerals, as well as mineral-specific tailored sets (TMINFF) improving per-mineral accuracy within the same functional form. Both GMINFF and TMINFF offer four levels of angle stiffness between all M-O triplets (O–M–O and M–O–M) using a single angle force constant of $k_a$ = 0, 250, 500, and 1500 kJ·mol$^{-1}$·rad$^{-2}$, so that coordination rigidity, vibrational response, and elastic behaviour can be tuned within a single functional form, with the lower and upper limits close to the angle stiffness of the CLAYFF ($k_a$ = 0 kJ·mol$^{-1}$·rad$^{-2}$) and INTERFACE ($k_a$ = 1422.56 kJ·mol$^{-1}$·rad$^{-2}$) force fields. In analogy with CLAYFF being derived from the SPC water model, MINFF is based on the oxygen-LJ parameters from the OPC3 water model,[28] with the goal of maximising compatibility with modern aqueous and ion-pair potentials including the recent 12-6-4 ion-pair potentials for polyvalent ions accommodating polarizability effects. MINFF is therefore a separate parameterisation rather than an interpolation between CLAYFF and INTERFACE, since the single angle term on all M–O triplets, the DDEC6-derived metal charges, the OPC3-derived oxygen Lennard-Jones parameters, and the paired general and tailored parameter sets fitted against more than 30 minerals together define its own balance between transferability and structural fidelity.

In parallel, the recent years have also seen rapid progress in alternative paradigms, reactive force fields tailored to earth materials,[29] several tailored or general machine-learning interatomic potentials (MLIP) such as MACE-MP, MTPs, and deep-neural-network potentials for silica, glasses, and zeolites,[30–32] and benchmark suites that emphasize transferability across observables (elastic constants, phonons, surface energies, defect formation) rather than only static structure. These approaches deliver impressive accuracy for specific chemistries but, in their current form, generally remain expensive for the long time- and length scales required to study large and fully hydrated systems, or do not handle substitutionally disordered clay systems well since they are typically trained on small ideal systems and data from DFT. Furthermore, they are not straightforwardly combined with established organic and biomolecular force fields that classical mineral force fields can provide. There is accordingly a continuing need to improve the classical, interpretable force fields that maintain the low-cost and high-compatibility advantages for various organic and inorganic solutes, and that can support chemically diverse mineral families as bulk or hydrated surfaces, with edge or substitution sites and other defect-type sites. The remainder of this paper describes how these design choices are realised and benchmarked.

MINFF is based on four core design elements: (i) a single and transferable angle term which depending on which MINFF parameter set uses $k_a$ = 0, 250, 500, or 1500 kJ·mol$^{-1}$·rad$^{-2}$, for all M–O triplets (M = Si, Al, Fe, Mg, Ca, Ti, Li, F), allowing tunable control over structural rigidity, vibrational response, and elastic behaviour, (ii) a set of M atom-type charges averaged from DDEC6 analysis,[33,34] of DFT-optimised reference structures, with oxygen charges assigned per-topology via the nearest-neighbour charge-smearing scheme of Lammers et al. (2017),[35] allowing for an extendable set of O atom-types, (iii) a chemically resolved atom-typing scheme implemented in the *atom* MATLAB and more recent *atomipy* Python libraries,[9,36,37] distinguishing tetrahedral and octahedral metal sites, basal, apical, and edge oxygens, and substitutional centres, and (iv) Lennard-Jones $\sigma$ and $\epsilon$ parameters fitted by iterative non-linear least-squares optimisation against experimental unit-cell metrics, bond distances, and bond angles. This fitting procedure was performed in three steps. First, the GMINFF parameter set was optimised synchronously across a training set of more than 30 minerals, using oxygen Lennard-Jones parameters equal to the water-oxygen parameters in OPC3 by default to ensure compatibility with this modern water model and ion-pair models. In a second and third optimisation step, tailored parameter sets with independently optimised M atom-type (second optimisation step, denoted TMINFF-M) and then O atom-type (third optimisation step, denoted TMINFF-MO) parameters were optimised. In addition to this, all hydroxide groups in MINFF share a single O–H bond term for all parameter sets, with a bond distance of 0.09572 nm and charge $q_H$ of +0.4.

The training set for the MINFF parameter set consisted of more than 30 minerals, including phyllosilicates (Kaolinite, Pyrophyllite, Talc, Montmorillonite, Nontronite, Dickite, Nacrite, Hectorite), tectosilicates (Quartz, Cristobalite, Coesite), simple oxides (Corundum, Hematite, Maghemite, Periclase, Lime, Wüstite, $Li_2O$, Anatase, Rutile, Akdalaite), oxyhydroxides (Boehmite, Diaspore, Goethite, Lepidocrocite), hydroxides (Brucite, Gibbsite, Portlandite), an Imogolite-type nanotube precursor (Imogolite), an orthosilicate (Forsterite), and a fluoride (Fluorite, $CaF_2$). Since the mineral-specific and tailored MINFF (TMINFF) parameters were also optimised individually for each of these structures, this provides a direct test of how much accuracy is gained by sacrificing strict transferability by tailoring either the M (second optimisation) or both the O atom-type (third optimisation) Lennard-Jones parameters. Across the full benchmark library, MINFF reproduces unit-cell parameters with median deviations of ca. 0.6% or below, atom-type distances within ca. 0.7–1.4%, and bond angles within ca. 0.1–0.4% when the angle term is active, while the Bond Valence Sum-derived GII decreases from ca. 0.23 to 0.11 as the angle term is stiffened, below the values obtained with CLAYFF and INTERFACE for the subset of minerals tested with both. All parameters, structure files, optimisation scripts, and topology generators are openly released through the MINFF GitHub repository[36] and through the online MINFF/CLAYFF topology server at www.atomipy.io,[37] which produces ready-to-run GROMACS, LAMMPS, and NAMD topology files.

The remainder of this paper is organised as follows. Section 2 describes the functional form of MINFF, the atom-typing scheme, and the partial-charge assignment strategy. Section 3 details the optimisation workflow, the choice of reference structures, the angle-force-constant parameter sets, and the validation protocol. Section 4 presents benchmark library results for unit-cell metrics, bond distances, bond angles, and Bond Valence Sum-derived stability across the full mineral library, with CLAYFF and INTERFACE included as reference points for a subset of their compatible minerals. Section 5 illustrates the practical consequences of these improvements through case studies of hydrated Montmorillonite systems, including ditrigonal cavity geometry, basal plane water density maps, basal spacings as a function of water content, and immersion energies. Section 6 extends the case study to the adsorption of an organic solute, comparing potential-of-mean-force profiles for zwitterionic Ciprofloxacin at the Na-Montmorillonite basal plane across the three force fields and both cavity geometries. Section 7 discusses the trade-off between general and tailored parameter sets, the role of the angle force constant, vibrational and elastic behaviour, and the interpretation of the dispersion energy in comparison to DFT-D3/D4. Section 8 summarises the strengths, current limitations, and ongoing extensions of MINFF, and provides a brief outlook on its role alongside other emerging force field methods in this field.

## 2. Force field formulation

### 2.1 Functional form

A central design choice in MINFF is that the M–O separation is not controlled by bond stretch terms but by the combination of (a) electrostatics, (b) Lennard-Jones repulsion/dispersion, and (c) the M–O–M and O–M–O angle terms acting on the surrounding O atom-types. This preserves the ability of MINFF to capture the directional rigidity of bonded force fields such as INTERFACE, yet offers a fully non-bonded character like in CLAYFF, and with it the ease of treating substitution and edge sites, or other exotic topology disorder.

Overall, MINFF is a class I fixed-charge force field, described below by the GROMACS-style Eq. 1–5, where ij or ijk indicate interacting atom pairs and triplets.

$$V_{Tot} = V_{Coul} + V_{LJ} + V_{angles} + V_{bonds,OH} \quad Eq. 1$$

The total potential energy $V_{Tot}$ (Eq. 1) comprises both non-bonded and bonded terms, where the former consists of a Coulombic (Eq. 2) and a 12-6 Lennard-Jones interaction term (Eq. 3) which uses the Lorentz–Berthelot mixing rules, meaning an arithmetic mean for $\sigma$ and geometric mean for $\varepsilon$.

$$V_{Coul}(r_{ij}) = \frac{q_i q_j}{4\pi\varepsilon_0\varepsilon_r r_{ij}} \quad Eq. 2$$

$$V_{LJ}(r_{ij}) = 4\varepsilon_{ij}\left[\left(\frac{\sigma_{ij}}{r_{ij}}\right)^{12} - \left(\frac{\sigma_{ij}}{r_{ij}}\right)^{6}\right] \quad Eq. 3$$

The angles are described by a single transferable harmonic angle potential (Eq. 4) on every M–O–M and O–M–O triplet ($0 \leq \theta_0 \leq 180$) involving the supported metal atom-types, including F (M ≡ Si, Al, Fe, Mg, Ca, Ti, Li, F):

$$V_{angles}(r_{ijk}) = \frac{1}{2}k_a(\theta_{ijk} - \theta_0)^2 \quad Eq. 4$$

In Eq. 4, $k_a$ is set to either 0, 250, 500, or 1500 kJ·mol-1·rad-2, depending on the specific MINFF parameter set, where the equilibrium angle $\theta_0$ for each individual $ijk$ triplet is taken from the initial reference structure. With $k_a$ = 0 the corresponding MINFF parameter set basically mimics CLAYFF, having fully unconstrained interactions apart from the O–H hydroxyl bonded terms described below. The contrary $k_a$ = 1500 kJ·mol-1·rad-2 parameter set mimics the angular stiffness of fully constrained INTERFACE which at least for silicates uses $k_a$ = 1422.56 kJ·mol-1·rad-2 in addition to pair-wise bonded terms.4 In practice, intermediate values of 250 and 500 kJ·mol-1·rad-2 are recommended, as they provide the best compromise between structural rigidity and vibrational response (Section 7).

For all MINFF parameter sets, the hydroxyl groups are described by a separate harmonic angle term for the M–O–H angle, with parameters adapted from the CLAYFF extension of Pouvreau et al. (2017, 2019),11,12 thus $\theta_0$ of 110 degrees and a $k_{a,MOH}$ of 125.52 kJ·mol-1·rad-2 for all parameter sets. The corresponding O–H harmonic bond term was, however, set to $k_{b,OH}$ being 441050 kJ mol-1 nm-2 according to Eq. 5, optimised against the Kaolinite structure of Bish (1993),38 see Table 1 below.

$$V_{bonds,OH}(r_{ij}) = \frac{1}{2}k_b(r_{ij} - b_{OH})^2 \quad Eq. 5$$

Although MINFF was designed without bonded terms, an M-O bonded term using $k$b,MO = 251 040 kJ mol−1 nm−2 has been tested successfully on clay edge sites but is typically not recommended.

Since the Lennard-Jones parameters of the O atom-types in the GMINFF are taken from the three-site OPC3 water model, it is the recommended water model for MINFF by default, but SPC, SPC/E as well as TIP3P, and TIP4P-style models have been tested and remain compatible by virtue of the OPC3-derived oxygen Lennard-Jones parameters used throughout the metal–oxygen network (Section 2.4).

Figure 1 exemplifies the angle terms used for the M-O triplets as well as the angle and bond terms used for O-H hydroxyl groups in MINFF for a phyllosilicate TOT substructure, including the derivation of the different partial charges detailed below in Section 2.3.

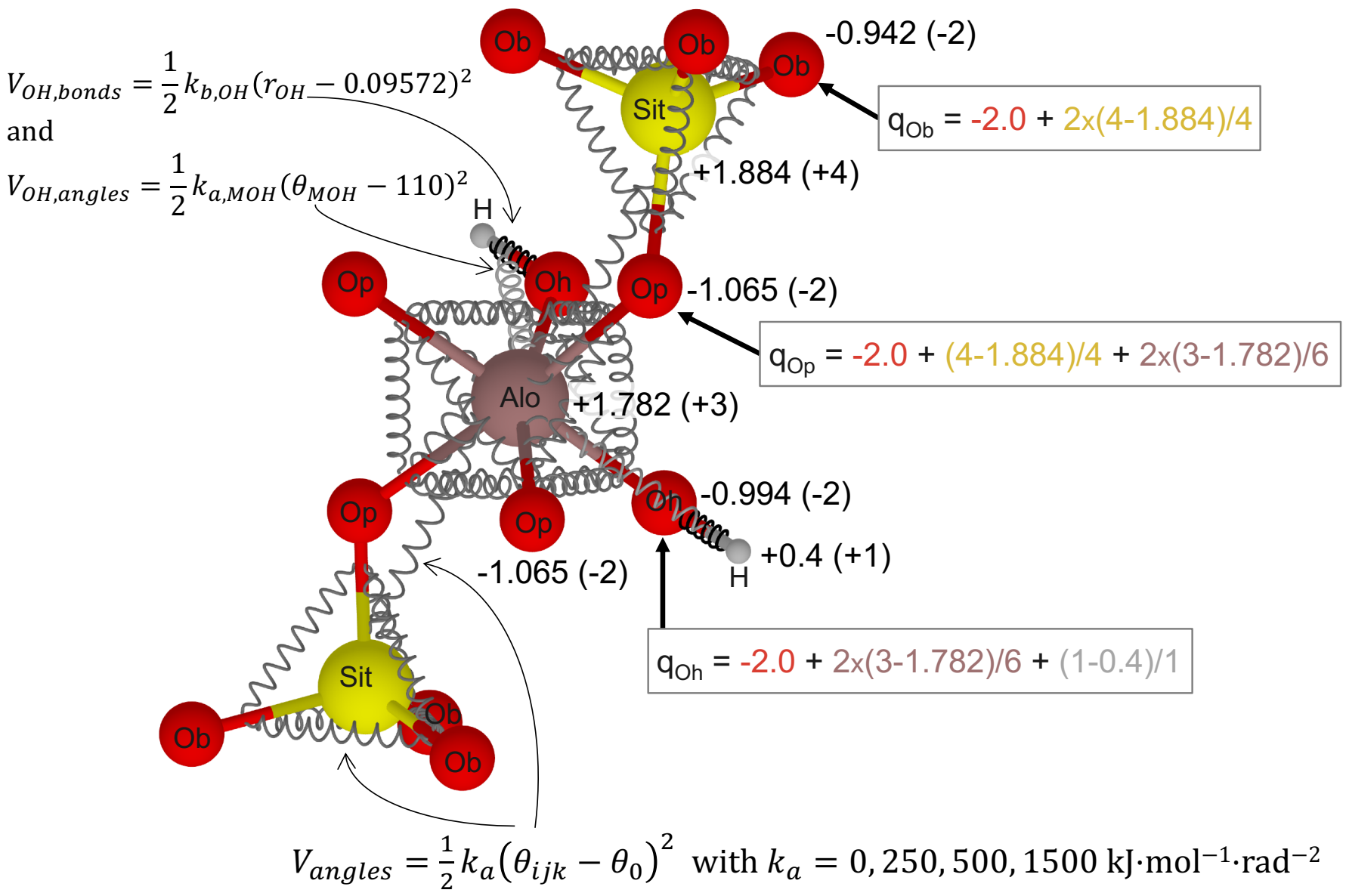

Figure 1. Example of the MINFF atom-types for a smectite TOT subunit with partial charges (denoted q) and formal charges (in parentheses). The boxes show the arithmetic of charge calculations for the basal (Ob), apical (Op) and hydroxyl (Oh) oxygens according to Eq. 6. Note that the sum term in Eq. 6 covers all coordinating neighbours to each oxygen, including atoms not shown in the figure.

## 2.2 Atom typing

MINFF uses a site-specific atom-typing scheme implemented in the *atom* (MATLAB/Octave) and *atomipy* (Python) libraries, where atom-types are assigned based upon nearest-neighbours, using cutoffs of 0.245 nm of each M or O site, and within 0.12 nm for any H. For heavily strained unit cells, however, these cutoffs may need to be changed for the atom-typing scheme to work. Overall, the scheme distinguishes:

(i) **tetrahedral cations** (Sit, Alt, Fet) and **octahedral cations** (Alo, Mgo/Mgh, Feo(2/3), Cao/Cah, Tio, Lio). Separate types are used for substitutional Mg and Fe in dioctahedral sheets and for Al or Fe(III) substituting Si in tetrahedral sheets. For Mg, the assignment between the oxide-like Mgo and the hydroxide-like Mgh type is composition-aware: Mgo is used where Mg sits in an oxide framework (Forsterite, Periclase) or in an Al-dominated dioctahedral sheet bearing only minor isomorphic Mg(II)→Al(III) substitution (mica, smectite), while Mgh is used in Brucite and in trioctahedral phyllosilicate sheets where every octahedral site is filled and the local ligand field is Brucite-like (Talc, Hectorite). The same oxide-vs-hydroxide split is applied to calcium: Cao is reserved for Ca in a pure oxide environment (Lime, CaO), whereas Cah is used in Portlandite $Ca(OH)_2$ and in Fluorite $CaF_2$, both of which present Ca with a hydroxide- or halide-like first coordination shell rather than oxide ions. In addition to the tetrahedral and octahedral M sites, from equation 6 defined below, it is in principle also possible to estimate the corresponding partial charge of defect and undercoordinated sites. Hence, although not as rigorously tested, MINFF also includes a few undercoordinated edge-atom-types, labelled Mge, Ale, Fee2, Fee3.

(ii) **oxygen environments**: where the main O atom-types describe and differentiate between bridging basal oxygens (Ob), apical oxygens at the tetrahedral–octahedral interface (Op), and structural hydroxyls (Oh). Due to the charge-smearing scheme, the list of O atom-types can also be extended to include arbitrary edge (Oe) or undercoordinated oxygens that arise when terminating a clay edge or an oxide surface, or atom-types that coordinate at isomorphic substitution sites, thereby capturing the local charge perturbations on the neighbouring oxygens. By default, the capitalised MINFF O atom-type labels distinguish basal, apical, and hydroxyl oxygens, as well as oxygens coordinating substitution sites or undercoordinated M sites. For multi-element systems with many defect or oxidation states, a custom naming scheme may be preferable, so that sites with different partial charges do not share the same atom-type labels.

(iii) **hydroxyl hydrogens** (H), carrying a single transferable hydroxyl hydrogen atom-type. An optional He label can be used to distinguish edge hydrogens from bulk lattice hydrogens.

Overall, these atom-typing criteria have been tested for over 60 mineral types, including classes of silicates (including many zeolites) and other mineral types, not part of the original training set. In case of ambiguity for unlisted mineral lattices, a short MD test run observing the unit cell metrics would often reveal the most suitable atom-types.

### 2.3 Partial charges

In MINFF, all the partial charges are shared between the different parameter sets of both the general and tailored parameter sets and across all minerals, assuring their transferability across structures as a key sanity check of the parameterisation (Section 4).

The partial charges in MINFF were derived from DDEC6 charge analysis,[33,34] obtained from CP2K geometry optimisation calculations of the reference mineral structures, including Montmorillonite with octahedral and tetrahedral substitutions based on configurations reported by Fonseca et al. (2017).[39] The DDEC6 charges were computed from electron densities mapped onto four multigrids with a plane-wave energy and relative cutoff of 400 Ry and 60 Ry, using the GPW formalism, the PBE functional, DFT-D3 dispersion correction, GTH pseudopotentials, DZVP-MOLOPT-SR-GTH basis sets, and an SCF convergence set to $1\times10^{-6}$ Hartree.

Overall, the resulting M atom-type charges are broadly similar to those in CLAYFF and substantially higher than those in INTERFACE. Most M atom-types carry slightly larger and more mutually consistent values than their CLAYFF counterparts. The main exception is tetrahedral Si (Sit), assigned +1.884 in MINFF, compared with +2.1 in CLAYFF (+1.1 in INTERFACE).

Across the tested minerals, the computed charges for Si, Al, Mg, Ca, Ti, Li, and F were rather consistent, with standard deviations typically below 5% of the mean of the elements mentioned, and 11% for H, due to its lower absolute charge, set to +0.4 in MINFF (average DDEC6 charge was +0.399). The tabulated M atom-type charges were rounded (within 2% of the respective DDEC6 averages) to reduce the number of significant figures in the resulting oxygen charges described in detail below. For Mg and Ca, this general level of consistency was only possible if allowing for two separate sets of atom-types, hence the use of Mgo and Mgh, as well as Cao and Cah, respectively. For Fe, the charges were based on spin-polarised DFT geometry optimisation, including a Hubbard correction of 1.9 eV for Fe-bearing smectites as suggested by Kéri et al. (2019).[40] The charges were calculated for both common Fe-oxides as well as Fe(II) and Fe(III) lattice sites in smectites, covering different substitution environments (single Fe, Fe-Mg or Fe-Fe pairs). Spin-polarisation DFT parameterisation is provided in the Supporting Information. Because the DDEC6 charges of the tested Fe minerals varied appreciably across oxidation states, mineral types, and coordination environments, we assigned fixed values of +1.184 and +1.5 to $Fe^{2+}$ and $Fe^{3+}$, respectively, which are broadly consistent with the DDEC6 distribution within each oxidation state.

In CLAYFF, M atom-type charges were set to minimise the number of both O atom-types, so that unsubstituted minerals like Quartz, Corundum, Kaolinite, and Brucite, etc. only need two O atom-types (one bridging and one hydroxyl). The idea behind MINFF is, however, to make it extendable and accommodate an arbitrary number of coordination environments and hence O atom-types. To this end, the partial charge of each O atom-type in MINFF is computed from the local bonding topology using the nearest-neighbour charge-smearing equation adapted from Lammers et al. (2017),[35] see Eq. 6.

$$q_O = -2.0 + \sum_j \frac{(Z_{F,j} - Z_j)}{CN_j} \quad Eq.6$$

In Eq. 6 the partial charge of every O atom-type $q_O$ is obtained from the formal charge of O (-2.0) plus the sum of all the $j$ coordinating M (and H) neighbours' difference in their respective formal ($Z_F$) and partial charge ($Z$), divided by their respective coordination number, $CN$. A similar derivation can be found in Ilinca (2022).[41] The sum term hence represents the electron density O shares with its coordinating M sites, including H. This ensures that the local charge is integer-based and naturally accommodates any form of defect sites due to isomorphic substitutions, different protonation states or uncoordinated edge terminations, without requiring a hand-curated table of per-site oxygen charges. Note that this charge scheme can also reproduce all O atom-type charges in CLAYFF, where it has been used to extend it with custom atom-types.[35,42–44] Eq. 6 is less useful for INTERFACE, where O atom-type charges are not strictly derived from local stoichiometry. Overall, the lower charge of Si makes the basal O atom-types (denoted Ob) softer compared to CLAYFF. The net effect of the slightly higher charge of the actual metal atom-types compared to CLAYFF is, however, that MINFF redistributes negative charge more toward the O sites that immediately coordinate any defect sites. For defect sites, Eq. 6 can also be inverted and used to calculate the partial charge of undercoordinated M sites, *i.e.,* calculating $Z_j$ for any M site with a different coordination number $CN$, such as the edge M atom-types.

### 2.4 Lennard-Jones parameters and water-model compatibility

For all O atom-types in the GMINFF parameter set, the Lennard-Jones $\sigma$ and $\epsilon$ were set to those of the OPC3 three-site water model. This single design choice had two practical consequences. First, MINFF is expected to perform well in other OPC3-hydrated systems and with any OPC3-compatible solute potentials, including organic force fields and ion-pair parameters, such as the 12-6-4 potentials derived by Li and Merz (2015).[45–50] The model was also validated against SPC/E and TIP3P water, yielding comparable structural results for hydrated systems. Second, fixing the O atom-type Lennard-Jones parameters removed them as fitting variables in the synchronous optimisation, which substantially reduced parameter degeneracy and avoided drift toward unphysical, water-model-specific oxygen parameters.

Metal Lennard-Jones parameters were obtained by iterative optimisation (Section 3). Interestingly, initial testing with different combinations of different $\varepsilon$ and $\sigma$ values for several minerals revealed an approximate $\varepsilon \propto \sigma^{-1/6}$ power-law relation, producing seemingly equally optimised fits. Hence, to start the optimisation with physically meaningful values, initial values for M $\sigma$ and $\epsilon$ were adapted from the C6 dispersion coefficients from DFT-D3,[51,52] on the same reference structures used for charge analysis. As a further check on the physical meaningfulness of the optimised parameters, the resulting site-decomposed dispersion ($-1/r^6$) energy curves were also compared to the corresponding DFT-D3/D4 C6 dispersion curves computed on a Pyrophyllite system.

In addition to the GMINFF parameter sets, the two tailored parameter sets first independently optimised the M atom-type Lennard-Jones parameters with O atom-type parameters fixed to OPC3 values (denoted TMINFF-M), and secondly individually optimised O atom-type parameters (denoted TMINFF-MO), using the OPC3 values as initial guesses and the M atom-type values from the TMINFF-M parameter set as the starting point. These tailored sets achieved slightly tighter agreement with the unit cell metrics and M–O distances, but not all reference structures (Section 4), as is discussed later in the text.

## 3. Optimisation workflow and validation protocol

### 3.1 Reference structures and screening

The general MINFF training set comprised more than 30 minerals, see minerals 1–32 in Table 1, representing the principal structural families of interest. The reference unit cells were taken from low-temperature, single-crystal X-ray and neutron-diffraction refinements compiled in the Inorganic Crystal Structure Database (ICSD) and the primary mineralogical literature. Minerals numbered 33–43 in the benchmark library were not part of the main training set and represent alternative Montmorillonite structures to the Lee and Guggenheim (1981) structure with different isomorphic substitution sites based on a structure from Tsipursky and Drits (1984),[53] as well as the Muscovite structure reported by Rothbauer (1971)[54] used in INTERFACE as the basis for Montmorillonite. Each candidate structure was screened by Bond Valence Sum (BVS) analysis using the BVparm2016 parameter set,[55] from which the Global Instability Index (GII), defined as the root-mean-square deviation of the cation valence sums from their nominal values, was evaluated. Only structures with $G_{II} < 0.20$ were retained as primary fitting targets, since structures with higher $G_{II}$ values are highly strained, rarely found in nature, and cannot serve as reliable reference structures.[56,57] Experimental rather than DFT-optimised structures were used throughout, because initial attempts with DFT-relaxed unit cells occasionally increased the $G_{II}$ relative to the experimental starting structures, rendering them less suitable as fitting targets. The additional structures 33–43 consisted of Montmorillonite structures (denoted MT) with additional types of isomorphic substitution, as well as Muscovite, and were only used as a validation set for the GMINFF, as well as for optimisation of the TMINFF parameter sets.

Table 1. Reference mineral structures used in the MINFF benchmark library, including additional Montmorillonite structures, with unit-cell parameters and experimental reference Global Instability Index (GII). Parameters sourced from single-crystal X-ray or neutron-diffraction refinements (see text). Imogolite nanotube structure based on Kaolinite, and individual lattice parameters are not applicable.

| No. | Mineral | ***a*** (Å) | ***b*** (Å) | ***c*** (Å) | ***α*** (°) | ***β*** (°) | ***γ*** (°) | $G_{II}$ | Ref. |
|---|---|---|---|---|---|---|---|---|---|
| 1 | Kaolinite | 5.154 | 8.942 | 7.391 | 91.93 | 105.05 | 89.80 | 0.049 | a |
| 2 | Pyrophyllite | 5.160 | 8.966 | 9.347 | 91.18 | 100.46 | 89.64 | 0.071 | b |
| 3 | Talc | 5.293 | 9.179 | 9.469 | 90.57 | 98.91 | 90.03 | 0.075 | c |
| 4 | Forsterite | 4.753 | 10.190 | 5.978 | 90.00 | 90.00 | 90.00 | 0.090 | d |
| 5 | Brucite | 3.150 | 3.150 | 4.770 | 90.00 | 90.00 | 120.00 | 0.003 | e |
| 6 | Corundum | 4.766 | 4.766 | 13.010 | 90.00 | 90.00 | 120.00 | 0.024 | f |
| 7 | Quartz | 4.912 | 4.912 | 5.404 | 90.00 | 90.00 | 90.00 | 0.124 | g |
| 8 | Gibbsite | 8.676 | 5.070 | 9.721 | 90.00 | 94.57 | 90.00 | 0.093 | h |
| 9 | $Li_2O$ | 4.619 | 4.619 | 4.619 | 90.00 | 90.00 | 90.00 | 0.079 | i |
| 10 | Coesite | 7.136 | 12.369 | 7.174 | 90.00 | 120.34 | 90.00 | 0.111 | j |
| 11 | Cristobalite | 4.972 | 4.972 | 6.922 | 90.00 | 90.00 | 90.00 | 0.165 | k |
| 12 | Maghemite | 8.347 | 8.347 | 8.347 | 90.00 | 90.00 | 90.00 | 0.138 | l |
| 13 | Akdalaite | 5.575 | 5.575 | 8.761 | 90.00 | 90.00 | 120.00 | 0.196 | m |
| 14 | Boehmite | 2.851 | 12.120 | 3.736 | 90.00 | 90.00 | 90.00 | 0.033 | n |
| 15 | Diaspore | 4.400 | 9.390 | 2.840 | 90.00 | 90.00 | 90.00 | 0.128 | o |
| 16 | Periclase | 4.203 | 4.203 | 4.203 | 90.00 | 90.00 | 90.00 | 0.011 | p |
| 17 | Goethite | 4.625 | 9.990 | 3.037 | 90.00 | 90.00 | 90.00 | 0.170 | q |
| 18 | Hematite | 5.035 | 5.035 | 13.747 | 90.00 | 90.00 | 120.00 | 0.025 | r |
| 19 | Lepidocrocite | 3.870 | 12.400 | 3.060 | 90.00 | 90.00 | 90.00 | 0.117 | s |
| 20 | Wüstite | 5.261 | 3.033 | 3.060 | 90.00 | 124.65 | 90.00 | 0.090 | t |
| 21 | Lime | 4.811 | 4.811 | 4.811 | 90.00 | 90.00 | 90.00 | 0.166 | i |
| 22 | Portlandite | 3.585 | 3.585 | 4.904 | 90.00 | 90.00 | 120.00 | 0.012 | u |
| 23 | Fluorite | 5.462 | 5.462 | 5.462 | 90.00 | 90.00 | 90.00 | 0.038 | v |
| 24 | Nontronite | 5.277 | 9.140 | 9.780 | 90.00 | 101.00 | 90.00 | 0.171 | w |
| 25 | Montmorillonite, MT | 5.160 | 8.966 | 9.347 | 91.18 | 100.46 | 89.64 | 0.122 | b |
| 26 | Dickite | 5.161 | 8.960 | 14.460 | 90.00 | 96.77 | 90.00 | 0.097 | x |
| 27 | Hectorite-F | 5.274 | 9.117 | 13.561 | 90.00 | 97.69 | 90.00 | 0.071 | y |
| 28 | Hectorite-H | 5.274 | 9.117 | 13.561 | 90.00 | 97.69 | 90.00 | 0.128 | y |
| 29 | Nacrite | 8.906 | 5.146 | 15.664 | 90.00 | 113.58 | 90.00 | 0.083 | z |
| 30 | Imogolite | - | - | - | - | - | - | 0.173 | aa |
| 31 | Anatase | 3.783 | 3.783 | 9.509 | 90.00 | 90.00 | 90.00 | 0.137 | ab |
| 32 | Rutile | 4.592 | 4.592 | 2.958 | 90.00 | 90.00 | 90.00 | 0.055 | ac |
| 33 | MT cis-Oct $Fe^{2+}$ cis | 5.180 | 8.980 | 10.050 | 90.00 | 101.40 | 90.00 | 0.109 | ad |
| 34 | MT cis-Oct $Fe^{2+}$ trans | 5.180 | 8.980 | 10.050 | 90.00 | 101.40 | 90.00 | 0.109 | ad |
| 35 | MT cis-Oct $Mg^{2+}$/$Fe^{3+}$ cis | 5.180 | 8.980 | 10.050 | 90.00 | 101.40 | 90.00 | 0.127 | ad |
| 36 | MT cis-Oct $Mg^{2+}$/$Fe^{3+}$ trans | 5.180 | 8.980 | 10.050 | 90.00 | 101.40 | 90.00 | 0.107 | ad |
| 37 | MT cis-Oct $Mg^{2+}$ cis/$Fe^{3+}$ | 5.180 | 8.980 | 10.050 | 90.00 | 101.40 | 90.00 | 0.128 | ad |
| 38 | MT cis-Oct $Mg^{2+}$ trans/$Fe^{3+}$ | 5.180 | 8.980 | 10.050 | 90.00 | 101.40 | 90.00 | 0.108 | ad |
| 39 | MT cis-Tet $Fe^{3+}$ | 5.180 | 8.980 | 10.050 | 90.00 | 101.40 | 90.00 | 0.127 | ad |
| 40 | MT trans-Oct $Fe^{2+}$ cis | 5.180 | 8.980 | 10.050 | 90.00 | 101.40 | 90.00 | 0.098 | ad |
| 41 | MT trans-Oct $Mg^{2+}$/$Fe^{3+}$ cis | 5.180 | 8.980 | 10.050 | 90.00 | 101.40 | 90.00 | 0.110 | ad |
| 42 | MT trans-Tet $Fe^{2+}$ | 5.180 | 8.980 | 10.050 | 90.00 | 101.40 | 90.00 | 0.097 | ad |
| 43 | Muscovite | 5.192 | 9.015 | 20.046 | 90.00 | 95.73 | 90.00 | 0.241 | ae |

In addition to the minerals in Table 1,[38,53,54,58–85] several zeolite structures from www.iza-structure.org/databases/ were also part of the validation set of the GMINFF.

### 3.2 Optimisation scheme

Force field optimisation was performed by iterative MD simulations in GROMACS controlled by an automated workflow in MATLAB based on functions from the *atom* library,[9] which acted as both the topology generator and the outer-loop driver. At each iteration, the current parameter vector $p_i = \{\sigma_i, \epsilon_i\}$ was mapped onto the native reference structure topology of every mineral in the training set containing atom-type $i$, which was then subjected to a 60 ps anisotropic NPT simulation at 1 bar and 298 K with a 1 fs timestep. The non-bonded interactions used a 1.2 nm cut-off with a Verlet scheme with a potential-shift modifier, particle-mesh Ewald electrostatics using a 0.12 nm grid with interpolation to the fourth order, and dispersion corrections to both the energy and the pressure. The last 40 ps was used to compute the time-averaged observables used to compute a weighted sum of squared residuals, $\chi^2(p,n)$, from the corresponding $n$ experimental reference minerals, according to Eq. 7:

$$\chi^2(p,n) = \sum_n \left(\Delta\boldsymbol{r} + {}^{\Delta\boldsymbol{\theta}}/_{200} + {}^{\Delta\boldsymbol{Cell}}/_{10}\right)^2 \quad Eq.7$$

The sum in Eq. 7 covered all minerals in the training set via their time-averaged $\Delta\boldsymbol{r}$, $\Delta\boldsymbol{\theta}$ and $\Delta\boldsymbol{Cell}$ vectors, representing the deviations in individual pair M–O distances, individual O–M–O (and M–O–M) angles, and super-cell metrics, as compared to the corresponding reference structures. To balance the influence of the three contributions, both the angle and the super-cell metrics terms were arbitrarily scaled down by a factor of 200 and 10, respectively. The parameter vector $p$ was updated using a bound-constrained nonlinear least-squares solver with a trust-region-reflective algorithm, until $\chi^2$ converged.

### 3.3 Validation protocol

For every optimised parameter set we report (a) the percentage deviations of the unit-cell metrics ($a, b, c, \alpha, \beta, \gamma$) relative to the experimental reference, (b) the median percentage errors of pairwise distances and angles, aggregated across all unique atom-type pairs and triplets, (c) the post-MD $G_{II}$ computed from BVS on the time-averaged structure, and (d) for selected minerals, theoretical XRD profiles, vibrational power spectra, basal plane density maps of interfacial water, and elastic constants from finite-strain perturbations of the relaxed unit cell. For comparison, the same battery of diagnostics was applied to a subset of compatible minerals using CLAYFF and INTERFACE with identical simulation protocols. As with MINFF, the CLAYFF topologies were generated by the atom MATLAB library, whereas the INTERFACE topologies were generated by the CHARMM-GUI Nanomaterial Modeler.[9,86,87]

## 4. Benchmark results

### 4.1 Unit-cell metrics and Global Instability Index

Table 2 summarises the percentage deviations of the unit-cell metrics from experimental reference structures, together with the final $G_{II}$, averaged across the full benchmark library of 43 minerals. Three observations stand out. First, every MINFF parameter set, including the $k_a$ = 0 control, reduces the mean lattice-parameter deviations relative to CLAYFF, from ca. 1.2–1.7% per axis down to ca. 0.6% or below, and simultaneously lowers GII from 0.34 to 0.11–0.23 depending on the parameter set (0.11–0.16 for the sets with a non-zero angle term). Second, the $k_a$ = 250–500 kJ·mol$^{-1}$·rad$^{-2}$ range gives the smallest combined error in lattice parameters and BVS-derived stability. Increasing $k_a$ to 1500 continues to lower $G_{II}$ but at the cost of slightly larger lattice deviations (along a and b for GMINFF, and along $c$ for TMINFF) and noticeably blue-shifted vibrational bands. Third, the tailored TMINFF-M parameter sets match or surpass the structural fidelity of INTERFACE on essentially every metric, with mean deviations of ca. 0.13–0.19% along a and b, ca. 0.4% along c, and GII of 0.12–0.16.

Table 2. Percentage deviations of unit-cell metrics from experimental references and post-MD, as well as the Global Instability Index (GII), averaged across the full MINFF benchmark library (>30 minerals) for the different parameter sets. Unit for $k_a$ is [kJ·mol$^{-1}$·rad$^{-2}$].

| **Force field** | $k_a$ | ***a*** (%) | ***b*** (%) | ***c*** (%) | ***α*** (%) | ***β*** (%) | ***γ*** (%) | **$G_{II}$** |
|---|---|---|---|---|---|---|---|---|
| GMINFF | 0 | 0.57 | 0.52 | 0.63 | 0.21 | 0.48 | 0.06 | 0.228 |
| | 250 | 0.49 | 0.46 | 0.47 | 0.16 | 0.35 | 0.02 | 0.155 |
| | 500 | 0.51 | 0.48 | 0.46 | 0.16 | 0.33 | 0.02 | 0.133 |
| | 1500 | 0.59 | 0.56 | 0.46 | 0.18 | 0.34 | 0.02 | 0.111 |
| TMINFF -M | 0 | 0.32 | 0.32 | 0.49 | 0.07 | 0.06 | 0.03 | 0.229 |
| | 250 | 0.13 | 0.18 | 0.37 | 0.07 | 0.06 | 0.02 | 0.158 |
| | 500 | 0.16 | 0.19 | 0.38 | 0.07 | 0.05 | 0.01 | 0.146 |
| | 1500 | 0.16 | 0.19 | 0.41 | 0.08 | 0.06 | 0.01 | 0.120 |
| TMINFF -MO | 0 | 0.17 | 0.15 | 0.37 | 0.07 | 0.08 | 0.03 | 0.234 |
| | 250 | 0.16 | 0.12 | 0.32 | 0.15 | 0.07 | 0.03 | 0.139 |
| | 500 | 0.13 | 0.09 | 0.31 | 0.15 | 0.21 | 0.01 | 0.133 |
| | 1500 | 0.13 | 0.07 | 0.33 | 0.15 | 0.20 | 0.01 | 0.119 |
| CLAYFF[a] | - | 1.17 | 1.29 | 1.68 | 0.32 | 0.52 | 0.05 | 0.343 |
| INTERFACE[a] | 1422.56 | 0.37 | 0.27 | 0.70 | 0.12 | 0.21 | 0.02 | 0.210 |

a) Results for CLAYFF and INTERFACE represent a subset of the MINFF benchmark library, see the Supporting Information for details.

### 4.2 Pairwise distances and bond angles

A force field can reproduce unit-cell metrics accurately while still distorting individual coordination polyhedra, so the pairwise distance and angle distributions provide an independent, more demanding test. Table 3 reports the median percentage errors of individual atom-type as well as pairwise atom-type distances and angles across the full benchmark library. The angle term plays a more decisive role here than for the lattice metrics: GMINFF reduces the median atom-type distance error from 2.84% (CLAYFF) to 1.17% at $k_a$ = 500 kJ·mol$^{-1}$·rad$^{-2}$ and 0.85% at $k_a$ = 1500 kJ·mol$^{-1}$·rad$^{-2}$, the latter marginally below INTERFACE (0.88%), with corresponding median atom-type angle errors of 0.20% and 0.11% (cf. 0.79% for CLAYFF and 0.32% for INTERFACE). The tailored TMINFF parameter sets push these numbers further, making them broadly consistent with the precision of the experimental refinements themselves and on par with the more constrained INTERFACE force field.

Table 3. Percentage errors of median per atom-type distances and angles, as well as pairwise atom-type distances, across the full MINFF benchmark library in comparison with a selected set of minerals simulated with CLAYFF and INTERFACE. Unit for $k_a$ is [kJ·mol$^{-1}$·rad$^{-2}$].

| Force field | $k_a$ | Atom-type distance Err (%) | Atom-type angle Err (%) | Atom-pair distance Err (%) |
|---|---|---|---|---|
| GMINFF | 0 | 1.40 | 0.27 | 1.89 |
| | 250 | 1.36 | 0.25 | 1.75 |
| | 500 | 1.17 | 0.20 | 1.41 |
| | 1500 | 0.85 | 0.11 | 0.83 |
| TMINFF -M | 0 | 1.82 | 0.50 | 2.19 |
| | 250 | 1.27 | 0.30 | 1.61 |
| | 500 | 1.09 | 0.22 | 1.35 |
| | 1500 | 0.72 | 0.12 | 0.82 |
| TMINFF -MO | 0 | 1.60 | 0.46 | 1.77 |
| | 250 | 0.88 | 0.37 | 1.22 |
| | 500 | 0.74 | 0.26 | 0.98 |
| | 1500 | 0.72 | 0.14 | 0.84 |
| CLAYFF[a] | - | 2.84 | 0.79 | 3.11 |
| INTERFACE[a] | 1422.56 | 0.88 | 0.32 | 0.70 |

a) Results for CLAYFF and INTERFACE only represent a subset of the MINFF benchmark library, see the Supporting Information for details.

Aggregated over the full set of structures, the GMINFF parameters give a more accurate unit cell and a lower GII than CLAYFF for ca. 2/3 of the minerals tested, while TMINFF is the most accurate parameter set for >80% of the benchmark. For the few minerals where INTERFACE retains a marginal advantage in lattice metrics (notably MgO, CaO, and the well-constrained α-$Al_2O_3$ family), MINFF still delivers comparable $G_{II}$ values, together with the substitutional, edge, and zeolite atom typing.

### 4.3 Elastic constants and mechanical response

In order to see the effect of the angle terms in MINFF as compared to CLAYFF and INTERFACE, bulk moduli $K$ for selected minerals and full elastic constants ($C_{ij}$) were computed by finite-strain perturbation of the NPT-equilibrated unit cell at 298 K, following the protocol described in Section 3.3. Although reference elastic data are sparse or vary widely for the tested minerals, the relative trends across the different force fields indicate differences in the properties of the tested parameter sets. Results for a representative set of minerals (Kaolinite, Muscovite, Quartz, Cristobalite, Corundum, Brucite, Hematite, Lime, Periclase, Portlandite, Wüstite) are shown in Figure 2, alongside compiled experimental reference values.[88] Figure 2a compares the bulk modulus $K$ of eleven minerals computed with MINFF (for $k_a$ = 0–1500 kJ·mol$^{-1}$·rad$^{-2}$), CLAYFF and INTERFACE against experiment.

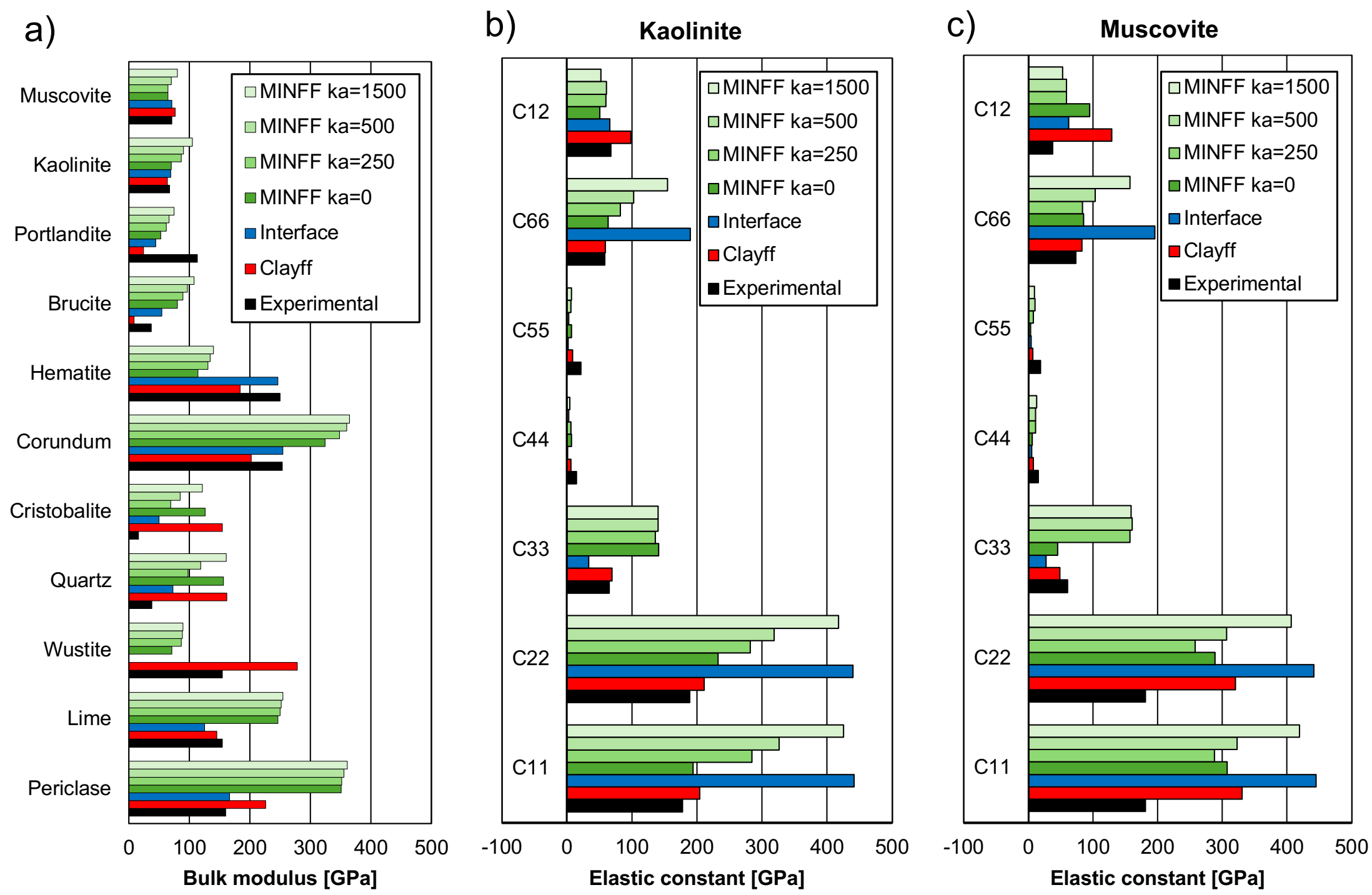

Figure 2. a) Bulk modulus K (GPa) for selected minerals and the main elastic constants Cij (GPa) for the phyllosilicates b) Kaolinite and c) Muscovite, computed by finite-strain perturbation of the NPT-equilibrated unit cell. MINFF values are reported for $k_a$ = 0, 250, 500, and 1500 kJ·mol-1·rad-2. CLAYFF and INTERFACE values were obtained using the same protocol.

For the sheet silicates (Muscovite, Kaolinite), all three force fields reproduce $K$ well and agree closely with one another, where $K$ for MINFF increases slightly with increasing $k_a$. For the dense, close-packed oxides, the updated parameters of the INTERFACE force field family are the most accurate for the oxide elasticity, whereas MINFF, with its larger charges, overestimates the bulk modulus of the main-group oxides (Periclase, MgO, Lime, CaO, Corundum, $Al_2O_3$) while also underestimating the iron oxides, the latter reflecting the difficulty of assigning fixed charges to a covalent transition-metal–oxygen bond. Interestingly, the framework silicates (Quartz, Cristobalite) are overestimated by all three force fields, as expected for rigid-ion potentials that cannot reproduce the tetrahedral-tilt compression mechanism responsible for the anomalously low experimental values. Notably, $K$ is relatively insensitive to $k_a$ for every structure type except these frameworks, supporting the use of a single transferable angle constant.

The diagonal and off-diagonal elastic constants for Kaolinite (Figure 2b) and Muscovite (Figure 2c) reveal the pronounced anisotropy of the phyllosilicates: the in-plane normal and shear constants ($C_{11}$, $C_{22}$, $C_{66}$) are stiff, whereas the layer-normal ($C_{33}$) and out-of-plane shear constants ($C_{44}$, $C_{55}$) are soft. For these phyllosilicates, INTERFACE systematically overestimates the in-plane constants due to rigid silicate bonds and angles, whereas CLAYFF, which lacks bonded terms, underestimates the bending- and shear-derived constants. Hence, at least for the phyllosilicates, MINFF with intermediate angle stiffness reproduces both the stiff in-plane and the soft shear and interlayer constants, yielding the smallest overall deviation from experiment and the best balance of transferability across structure types.

### 4.4 Angle stress analysis

To further quantify how the angle term controls coordination geometry across the tested minerals, a per-mineral angle strain factor $\langle(\theta - \theta_0)^2\rangle$ was defined from the angle term potential energy component, as the mean squared angular deviation from equilibrium, obtained from the angle term potential energy, normalised by $2k_a$ and the number of angle terms (Eq. 8), with the $k_a$-independent hydroxyl H–O–M contribution removed using the $k_a$ = 0 reference.

$$\langle(\theta - \theta_0)^2\rangle = 2V_{angles}/\left(k_a N_{angles}\right) \quad Eq. 8$$

Interestingly, the angle strain, tabulated in Table S3, was found to be ordered almost entirely by structural class, where dense close-packed oxides/halides (Lime, CaO, Periclase (MgO), Wüstite (FeO), $Li_2O$, Fluorite ($CaF_2$), Rutile and Anatase ($TiO_2$), Corundum ($Al_2O_3$) and Hematite ($Fe_2O_3$)), as well as the orthosilicate Forsterite have the lowest strain at 250 kJ·mol$^{-1}$·rad$^{-2}$ (ca. 0.0007–0.0013 rad$^2$), on average followed by the hydroxides, and the framework and sheet silicates, oxyhydroxides, and substituted octahedral models form a higher band (ca. 0.0035–0.0044 rad$^2$), with the mixed-valence/vacancy-ordered Maghemite having the highest strain (ca. 0.0094 rad$^2$). Overall, $k_a$ was found to monotonically reduce the strain for every mineral by a relative strain factor of 2.5–6.8 over the sixfold $k_a$ increase from 250 to 1500 kJ·mol$^{-1}$·rad$^{-2}$, confirming that, apart from the simple binary oxides, a single transferable angle term significantly improves geometries compared to their reference $\theta_0$. See summary of the strain values in Table S3. These trends parallel the elastic coefficients (Section 4.3) in which the framework silicates are likewise the most sensitive to $k_a$.

### 4.5 Vibrational power spectra of Kaolinite

To further probe the curvature of the potential energy surface directly on a silicate framework, power spectra of Kaolinite were computed to provide a qualitative comparison of the band positions across different MINFF force fields. This was accomplished by computing the velocity autocorrelation function (VACF) from 60 ps trajectories propagated with a 0.5 fs timestep, with velocities saved every 2 fs, and averaging spectra from 5 trajectories. The O–H force constant $k_b$ in MINFF was in fact optimised to Kaolinite and set to 441050 kJ·mol$^{-1}$·nm$^{-2}$ using a bond distance of 0.09572 nm, which places the hydroxyl bands between ca. 3620 and 3700 cm$^{-1}$, covering the inner and inner-surface bands reported for well-ordered KGa-1 Kaolinite,[89] whereas CLAYFF and INTERFACE use corresponding bond parameters producing bands being either blue- or red-shifted, as can be seen in Figure 3. At lower wavenumbers in the librational and lattice-vibration regions, the situation is reversed. For instance, the non-bonded CLAYFF Kaolinite lattice shows a single broad band between ca. 400 and 800 cm$^{-1}$ rather than discrete modes, which is the spectral counterpart of the lower bending- and shear-derived constants C44, C55, and C66 obtained with that model, whereas INTERFACE resolves the Si–O features sharply but blue-shifted towards higher wavenumbers, in line with its stiffer in-plane constants and scaled bond distances (+5%). In MINFF, the bands in this window shift upwards with increasing $k_a$ and the intermediate values correspond most closely to the reported KGa-1 Kaolinite bands at 792, 753, 698, 539, 471, and 431 cm$^{-1}$, whereas $k_a$ = 0 kJ·mol$^{-1}$·rad$^{-2}$ leaves this region unresolved, similar to CLAYFF, as expected when the angle terms are switched off. Similar to INTERFACE, however, with $k_a$ = 1500 kJ·mol$^{-1}$·rad$^{-2,}$ MINFF also produces additional blue-shifted bands not observed for the other MINFF or CLAYFF parameter sets.

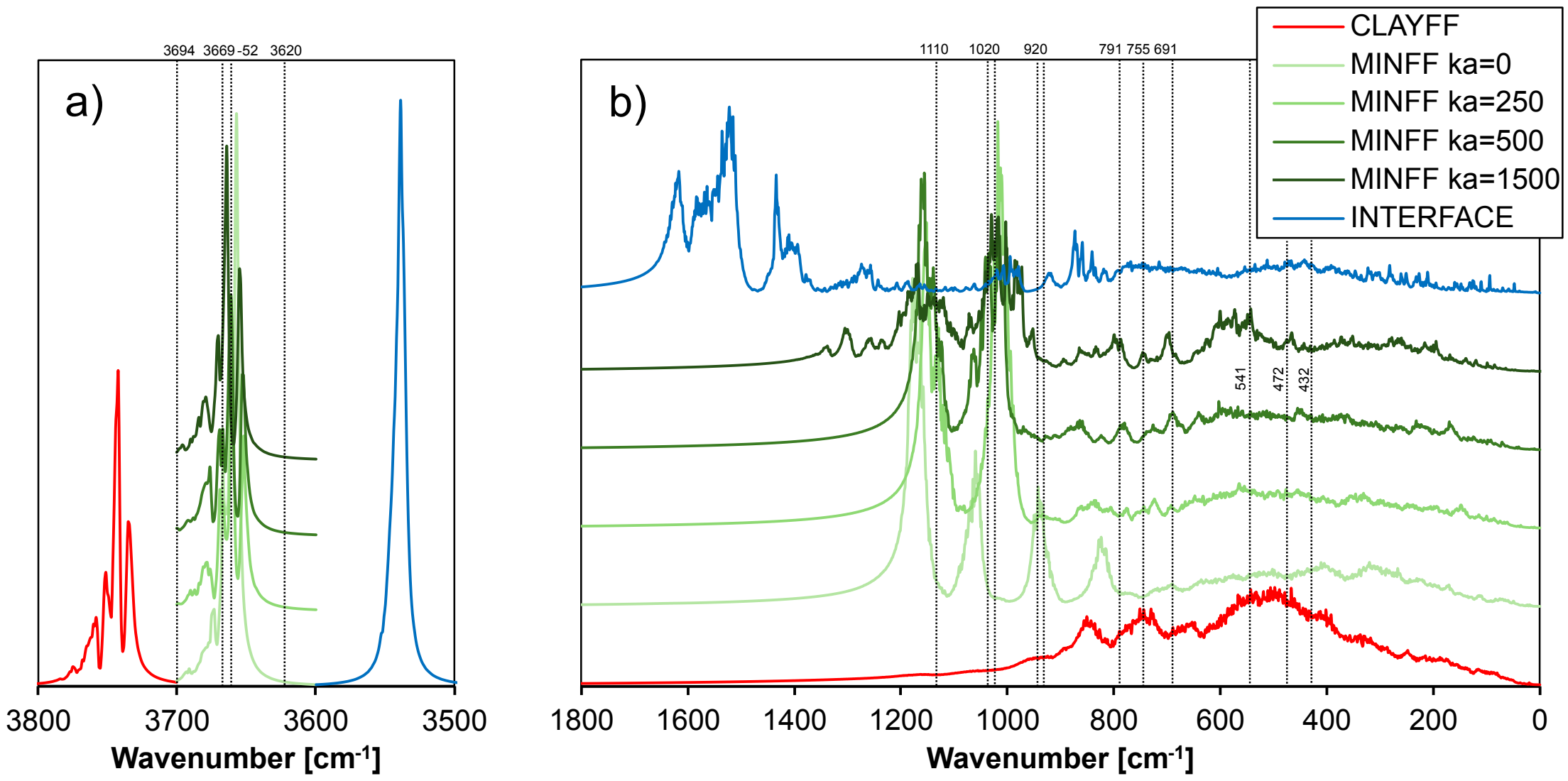

Figure 3. Power spectra comparison of Kaolinite computed from the VACF for CLAYFF, the GMINFF versus $k_a$, and the INTERFACE force field. a) The H-stretch region. b) The librational and lattice vibrations. Vertical lines indicate approximate vibrational band positions in cm$^{-1}$ for well-ordered Kaolinite KGa-1.89

### 4.6 Diffraction-based structural validation

To scrutinise the structural agreement between the general and tailored parameter sets of MINFF and CLAYFF and INTERFACE, comparisons of diffraction patterns are particularly useful, since XRD is the principal technique used to characterise mineral structures. Furthermore, in addition to comparisons of unit-cell metrics, pairwise distances, and bond angles, a diffraction pattern reflects the full atomic arrangement and thus tests the simulated structure as a whole against its reference structure. Accordingly, both the experimental reference structures and structures from unconstrained, fully anisotropic MD simulations with triclinic symmetry were compared. The diffraction patterns were calculated with the xrd_atom function of the atom MATLAB library for random-mount samples, and the structural agreement was quantified as the intensity difference between the simulated and reference XRD diffractograms ($I_{\mathrm{sim}} - I_{\mathrm{ref}}$) using the weighted profile factor $R_{\mathrm{wp}}$ (in %). Overall, the obtained $R_{wp}$ values spanned a wide range, noticeably higher than those typically reported for accurate experimental Rietveld refinements, underscoring the value of direct diffraction-based benchmarking. Across the subset of minerals also tested with CLAYFF and INTERFACE, GMINFF showed on par or better agreement to the reference structures in 75% of cases relative to CLAYFF and in 50% of cases relative to INTERFACE structures generated by the CHARMM-GUI Nanomaterial Modeler. Notable cases where CLAYFF outperformed the GMINFF parameter set were for Kaolinite, for which CLAYFF was optimised, and the Kaolinite polymorphs Dickite, Nacrite, as well as for Periclase.

GMINFF also underperformed in comparison to the mineral-specific INTERFACE parameterisations for Quartz, Corundum, Periclase, Lime, Brucite, and Portlandite. For the TMINFF parameter sets, however, only Corundum, Brucite, and Portlandite remained better described by INTERFACE, while for all other tested minerals TMINFF matched or surpassed the agreement between the theoretical reference and simulated diffraction patterns obtained with both CLAYFF and the corresponding INTERFACE parameter sets.

As with CLAYFF and the original INTERFACE force field implementations, MINFF was primarily designed for layered silicates, in particular clay minerals such as Kaolinite and the Montmorillonite derived from Pyrophyllite by isomorphic substitution. These two minerals are particularly interesting for benchmarking purposes because both are built from tetrahedral (T) and octahedral (O) sheets stacked into layers which are held together only by weak non-electrostatic interactions: van der Waals forces between the TOT-TOT layers of Pyrophyllite, and additionally interlayer hydrogen bonds in the TO-TO structure of Kaolinite. The benchmark library for MINFF included the reference structures commonly used in the CLAYFF literature, Kaolinite from Bish (1993) and Pyrophyllite from Lee and Guggenheim (1981), closely related to the orthogonalised structures used by the CHARMM-GUI Nanomaterial Modeler for INTERFACE. Since CLAYFF was specifically optimised against the Bish (1993) Kaolinite structure, the corresponding GMINFF parameter set with $k_a$ = 0, which was trained jointly on more than 30 additional minerals, yielded slightly higher $R_{\mathrm{wp}}$ values (Figure 4a). Introducing an angle term ($k_a$ = 250, 500, or 1500 kJ·mol$^{-1}$·rad$^{-2}$) substantially improved the GMINFF $R_{\mathrm{wp}}$ values, clearly demonstrating the value of the added angle terms. The same trend held for the tailored sets, with TMINFF producing $R_{\mathrm{wp}}$ values significantly lower than those of both CLAYFF and GMINFF. For Pyrophyllite, both GMINFF and TMINFF parameter sets yielded lower $R_{\mathrm{wp}}$ values than CLAYFF and INTERFACE, with the exception of the TMINFF-M case at $k_a$= 0, which gave a surprisingly high $R_{\mathrm{wp}}$ of 52.6%. Further inspection revealed that this same set produced the lowest $G_{\mathrm{II}}$ of all Pyrophyllite simulations, illustrating the difficulty of reproducing both unit-cell metrics and the local coordination environment of the M sites simultaneously, and the preference for optimising pair distances and bond angles over unit cell metrics in Eq. 7. Compared with MINFF and CLAYFF, INTERFACE, which performed overall well for the simple oxides and hydroxides, gave higher $R_{\mathrm{wp}}$ values for the two phyllosilicates, plausibly because it was parameterised against other orthogonalised reference structures accessible via the CHARMM-GUI Nanomaterial Modeler.

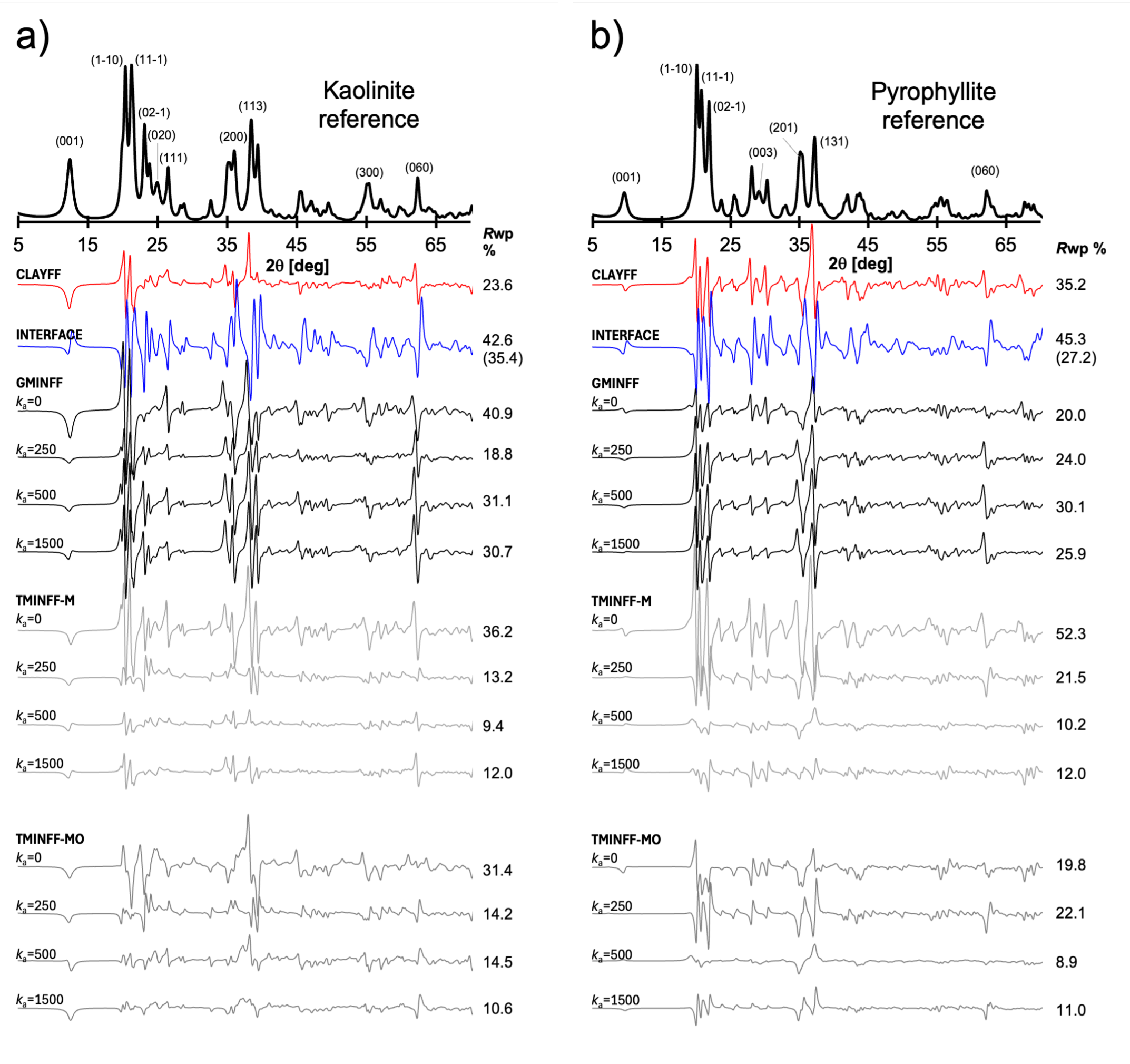

Figure 4. Theoretical XRD diffractograms calculated from the experimental reference structures of a) Kaolinite and b) Pyrophyllite, and the corresponding intensity difference from fully unconstrained and anisotropic MD simulations using CLAYFF (red), INTERFACE (blue), the GMINFF (black/gray) as well as the tailored parameter sets TMINFF-M and TMINFF-MO. $R_{wp}$ values in parentheses for INTERFACE represent orthogonal structures from CHARMM-GUI.

## 5. Case study: hydrated Na-Montmorillonite

The hydration properties of smectite clay minerals such as Na-Montmorillonite comprise another critical test system for a mineral force field such as MINFF, since the physical observables, such as basal spacing ($d_{001}$), swelling immersion energy, and interlayer structure and self-diffusion of water and other interlayer solutes couple intimately to the geometry of the ditrigonal cavity and the local ordering of substitutional sites. For this reason, simulations were performed as a function of diminishing water content using Na-Montmorillonite with the formula $Na_{0.66}[Al_{3.34}Mg_{0.66}][Si_8]O_{20}(OH)_4$, here modelled as a 6×4×3 supercell with SPC/E water and ion-pair potentials from Joung and Cheatham (2008).[90] For CLAYFF and MINFF, the Montmorillonite structure was based on Lee and Guggenheim (1981), whereas Montmorillonite modelled with INTERFACE was obtained from the CHARMM-GUI server and based on the mica structure of Rothbauer (1971). For further details, see Holmboe and Bourg (2014) as well as the Supporting Information.[91]

### 5.1 Basal spacing, immersion energies, and interlayer dynamics

The basal spacing $d_{001}$ as a function of the number of interlayer water molecules in the crystalline swelling regime (0W→3W) is reproduced quantitatively by all the tested force fields and cavity geometries, with the characteristic plateaus at $d_{001}$ of ca. 1.24, 1.56, and 1.88 nm matching commonly reported XRD results for Na-Wyoming-type Montmorillonite having similar surface charge.[92] However, the corresponding immersion-energy profiles show clear differences. When hydrated to 1W or higher, the GMINFF agrees well with the hydration energy of the tested CLAYFF simulations, regardless of cavity geometry and water model, all showing a global minimum of the immersion energy tied for the 2W and 3W hydration states. For the INTERFACE force field, the 3W hydration state is the global minimum within the crystalline swelling regime. Below 1W, however, a clear difference between MINFF and CLAYFF is observed, with the latter also showing a strong dependence on the cavity geometry, as evidenced by the immersion energy difference between 0W and 3W, as indicated in Figure 5b. Under this regime, MINFF yields an immersion energy difference closer to INTERFACE, regardless of the cavity geometry or the water model (SPC/E or OPC3). The two CLAYFF cases display the lowest and highest overall immersion energies of the whole set, 53.4 J/g (63 J/g with OPC3, not shown) for the relaxed hexagonal cavity geometry and 95.9 J/g for the constrained ditrigonal cavity geometry. This is a significant difference from the same regime for the INTERFACE and different MINFF cases, all of which had immersion energies within 3.5 J/g. Closer inspection of the CLAYFF cases at the fully dehydrated level showed that with a fully relaxed hexagonal cavity geometry, the interlayer $Na^+$ cations were the most deeply buried of all cases, sitting on average 0.7 nm from the layer centre, compared with 1.0 nm for the constrained ditrigonal CLAYFF case. For GMINFF, by contrast, a hexagonal cavity pushed the interlayer $Na^+$ cations outwards relative to the ditrigonal case, from 0.8 to 0.9 nm from the layer centre. For INTERFACE, the interlayer cations were found 1.2 nm from the layer centre.

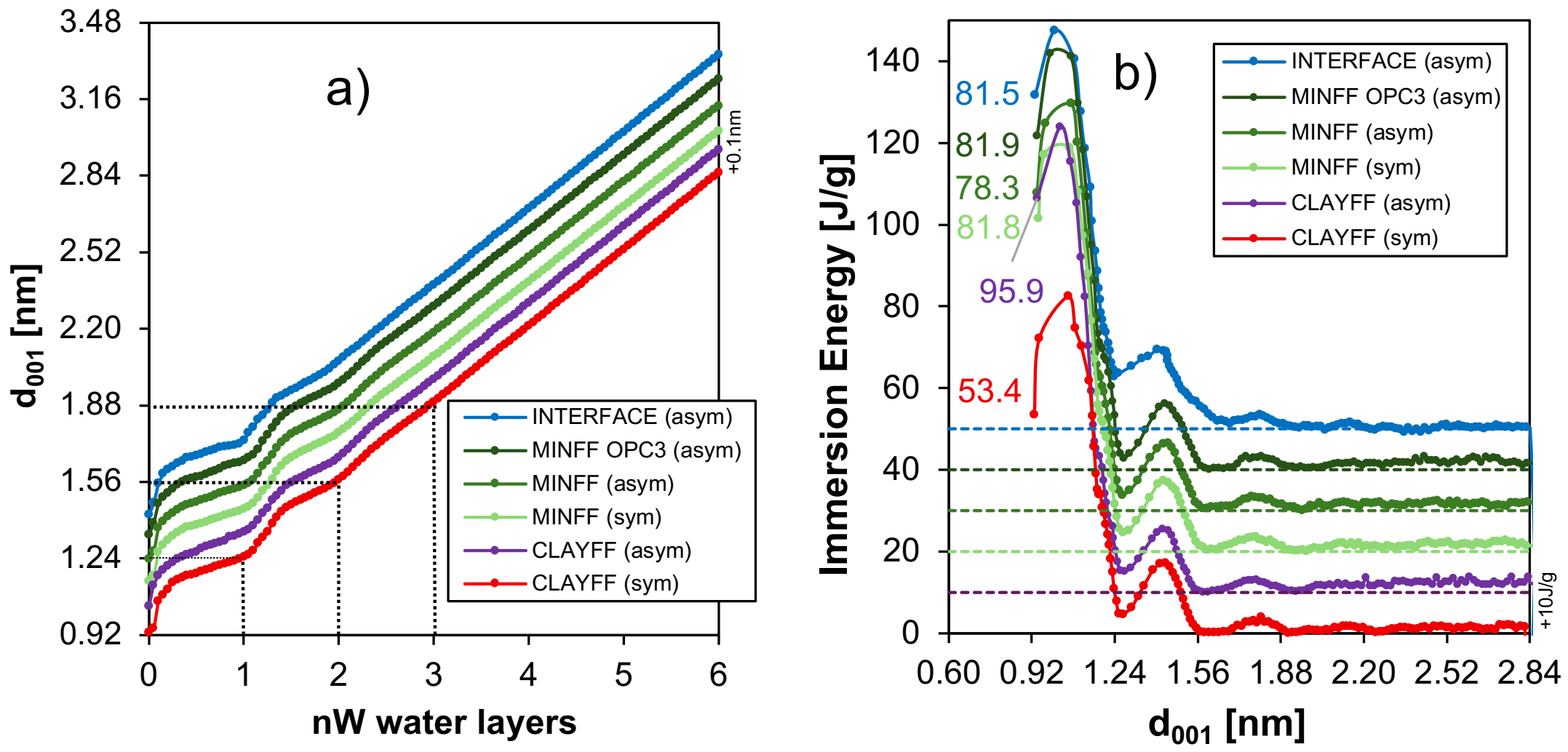

Figure 5. Hydration properties of Na-Montmorillonite, showing in a) the basal spacing (d001) dependence on the number of pseudo-monolayers of water (nW=1–6) taken as 5 water molecules per unit cell, for the different force fields and cavity geometry. b) The corresponding immersion energy in J/g. For clarity, the data are shown with an offset of +0.1 nm and +10 J/g in a) and b), respectively.

Self-diffusion coefficients of interlayer water were also computed from mean-square displacements, see Figure S3 in the Supporting Information. With ditrigonal cavities, the self-diffusion coefficients over the different hydration states for MINFF were ca. 4% lower compared to MINFF with hexagonal cavities, even at high hydration states. For CLAYFF, the same difference was up to ca. 20%, indicating that the atomic vibrations of the silicate framework, rather than the cavity geometry, play a greater role in interlayer water dynamics. INTERFACE using the same SPC/E water model and ion-pair potentials produced self-diffusion coefficients ca. 10% lower compared to the default CLAYFF (sym) and MINFF (asym) cases, which by themselves showed insignificant differences in diffusion coefficients, for both the SPC/E and OPC3 water models (Figure S3d).

**5.2 The effect of silicate sheet cavity geometry**

To investigate the effect of the basal oxygen cavity geometry on the intrinsic water organisation across the tested force fields, single-surface density maps of water oxygen and hydrogen from the 2W hydration state, with a $d_{001}$ of ca. 1.56 nm, were computed. Figure 6 shows the corresponding density maps for the GMINFF = 500 kJ·mol$^{-1}$·rad$^{-2}$, with either the native experimental ditrigonal basal plane cavities or the relaxed hexagonal symmetrical cavities. Corresponding data for CLAYFF and INTERFACE are shown in Figure S5 in the Supporting Information.

As can be seen in Figure 6, the lateral water density of the surface layer follows the structure imposed by the basal oxygen plane but is also affected by the dioctahedral sheet structure and the hydroxyl groups (see Figure S4) in Montmorillonite. With asymmetric and ditrigonal cavities, the water-oxygen density is organised into curved ridges that skirt the cavity rims and connect neighbouring cavities (Figure 6, top), at an angle determined by the hydroxyl groups in the octahedral sheet. The hexagonal and symmetric cavities instead produce more ordered water-oxygen (Ow) distributions with density maxima stretching over the cavity centres as two main corridors with a 60° angle, confirming the effect of the asymmetric ditrigonal cavities on surface water organisation.

Despite residing closer to the basal plane, the water-H atoms show more similar density distributions between the ditrigonal and hexagonal basal plane cavities, differing only slightly along the six ridges that connect neighbouring cavities. This is consistent with the nearest surface layer of water forming either a bidentate or monodentate surface-bound population, able to donate one hydrogen bond to the corner-sharing basal oxygens.[93] In the symmetrical case, four out of six ridges of water-H that do not coincide with the directions of the lattice hydroxyl groups in trans-vacant Montmorillonite (Figure S4) are equally pronounced, whereas over the ditrigonal cavities mainly the two ridges oriented next to the hydroxyl groups are most pronounced, producing a slightly more laterally anisotropic surface water-H distribution.

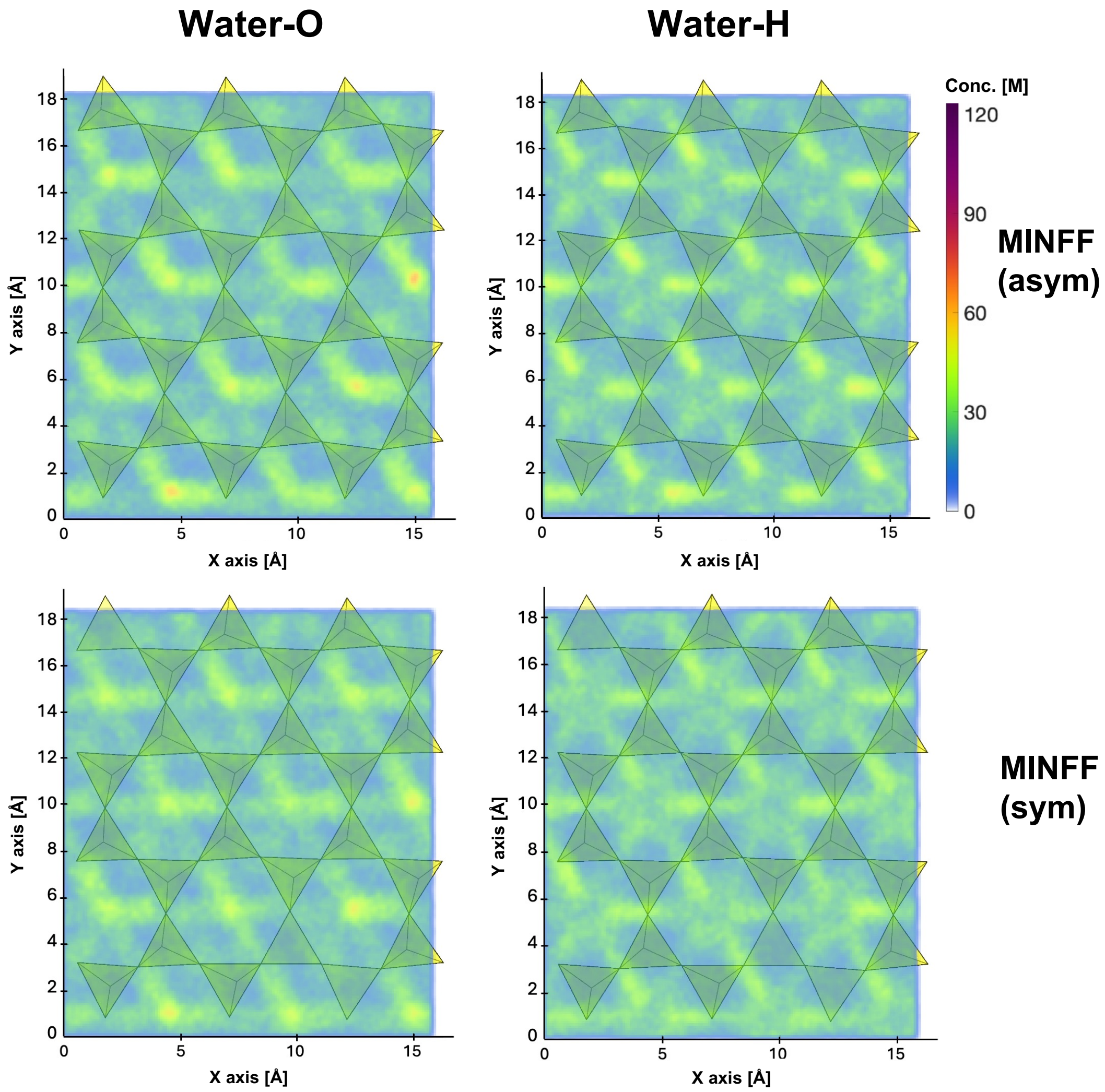

Figure 6. Density maps of water from half the 2W interlayer interfacing the Montmorillonite basal plane modelled with the GMINFF and $k_a$ = 500 kJ·mol-1·rad-2. Left column images show the density of water-oxygen and the right column the water-hydrogen density in molar concentrations. Top row images show the densities from the default MINFF having ditrigonal cavities (denoted asym), and the bottom row shows the corresponding system with relaxed hexagonal and symmetric cavities (denoted sym).

In comparison to CLAYFF and INTERFACE, GMINFF resembles the former force field to a much higher degree in terms of both the 1D and 2D density over the basal plane (Figure S5 and S6). Montmorillonite modelled with INTERFACE carries a larger tetrahedral rotation due to the Rothbauer (1971) structure, resulting in denser and point-like water-O maxima over the strongly ditrigonal cavities, producing a water-H density both closer to the basal plane with a higher degree of bidentate population, in line with the lower interlayer water self-diffusion (Figure S3c).

To test the significance of the slight difference in interlayer organisation, the 1D density profiles of the 1.56 nm 2W hydration state were converted to 1D XRD structure factors and used to calculate theoretical XRD patterns (CuKα). The calculations used a uniform lattice adapted from the Reynolds 2W smectite structure,[94] with the empirical CLAYFF-Mod with SPC/E interlayer model as a reference for the interlayer solutes,[18] which shifts the interlayer species away from the basal plane because its basal O atom-types are larger, which increases the *002* and *003* reflection intensity of smectites. Although the differences in the 1D density profiles are small, especially between the GMINFF and CLAYFF variants, the resulting step-normalised squared residuals (Rp) of the theoretical XRD profiles (Figure S7) rank the different systems as GMINFF (asym) + OPC3 (0.005) > CLAYFF (sym) + SPC/E (0.010) > MINFF (sym) + SPC/E (0.012) > MINFF (asym) + SPC/E (0.014) > INTERFACE (asym) + SPC/E (0.069), indicating that the OPC3 water model partially and seemingly induces the same effect as CLAYFF-Mod.

## 6. Case study: Drug molecule adsorption to Na-Montmorillonite

To probe the effect of the GMINFF, CLAYFF, and INTERFACE force fields and the effect of the ditrigonal cavity geometry on organic molecule adsorption, potential of mean force (PMF) profiles representing the free energy of adsorption of the antibiotic Ciprofloxacin (zwitterionic at pH 7) to Na-Montmorillonite were also computed, following an established umbrella sampling workflow.[43] For these simulations, TIP3P was used as the common-ground water model and Ciprofloxacin was described by CGENFF. Although all MINFF parameter sets were optimised with dispersion correction for both energy and pressure (same as for GAFF), turning the dispersion correction off in order to use CGENFF had insignificant effects on the structural data of the clay.

Also in this case study, two system topologies were built for CLAYFF and MINFF. In the first case, the asymmetric INTERFACE topology was converted into a MINFF topology and into a CLAYFF topology, both carrying the MINFF-type M–O–M angle terms ($k_a$= 500 $kJ\cdot mol^{-1}\cdot rad^{-2}$). In the second case, relaxed and symmetric hexagonal cavities (denoted sym) were used, generated by first running default unconstrained CLAYFF simulations, and also using an averaged CLAYFF structure as the basis for a MINFF topology with $k_a$= 500 $kJ\cdot mol^{-1}\cdot rad^{-2}$. Taken together, the two cases allowed for CLAYFF and MINFF to be compared while having similar cavities. The resulting PMF profiles obtained for INTERFACE, CLAYFF, and MINFF are shown in Figure 7, with the corresponding minima and sorption coefficients collected in Table 4.

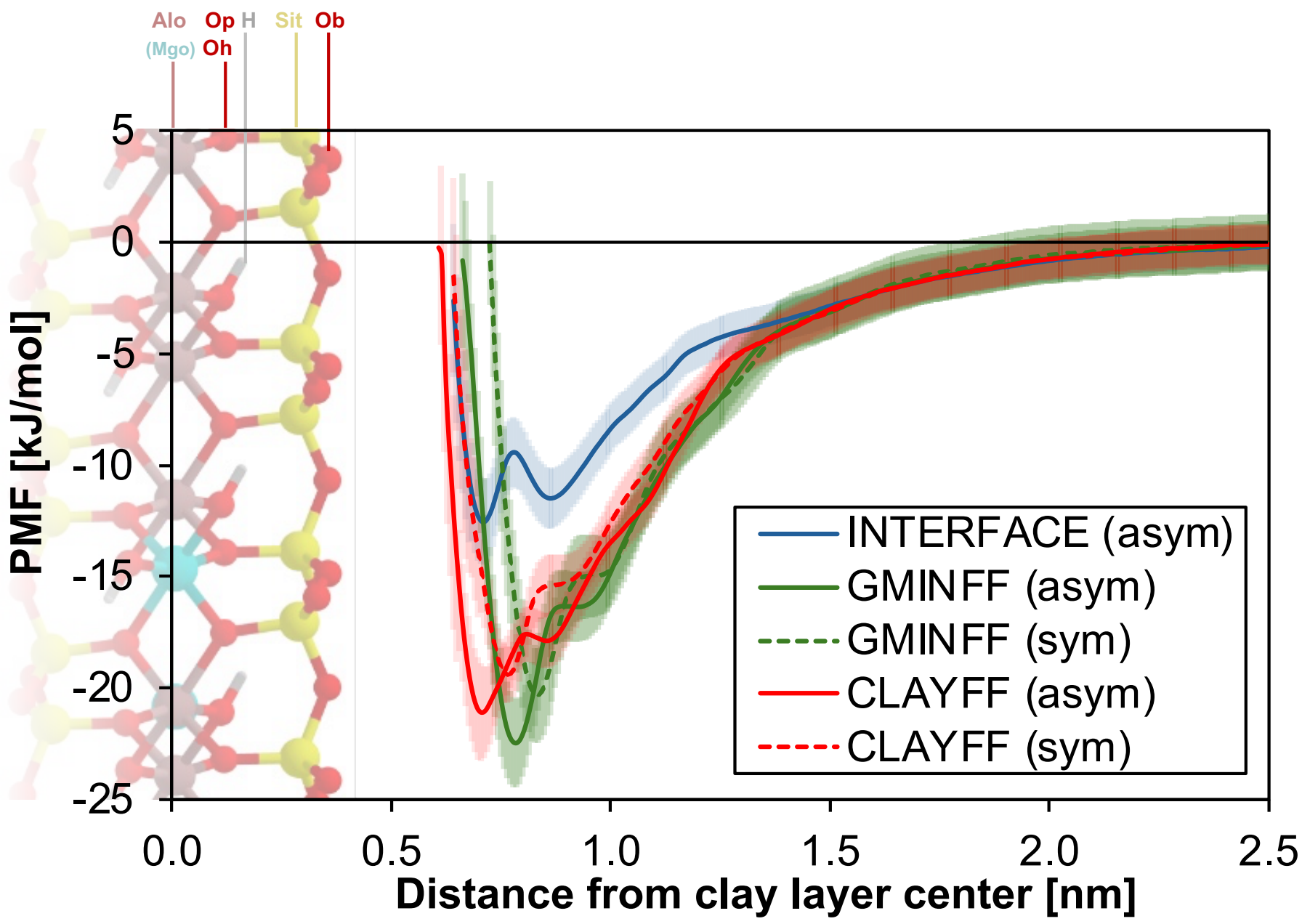

Figure 7. Potential of mean force profiles for the zwitterionic Ciprofloxacin molecule from the centre of a Na-Montmorillonite layer (indicated by the MINFF atom-types), modelled with INTERFACE, CLAYFF, and the GMINFF. Solid lines (asym) result from simulations with the experimentally realistic asymmetric ditrigonal cavities in the basal oxygen plane, enforced either by bond and angle constraints (INTERFACE) or by angle constraints alone ($k_a$= 500 kJ·mol$^{-1}$·rad$^{-2}$ for both CLAYFF and MINFF). Dashed lines (sym) result from simulations in which the basal oxygen cavities were symmetric and hexagonal.

Between the force fields, INTERFACE gives a distinctly shallower PMF profile and smaller free energy of adsorption than CLAYFF and MINFF, which follows from its smaller partial charges and from basal-oxygen Lennard-Jones ε values well below those of CLAYFF and MINFF, which are themselves comparable.

Table 4. Global minimum of the Ciprofloxacin PMF profiles, its distance from the centre of the Na-Montmorillonite layer, and the sorption coefficient Kd obtained by integrating each PMF profile as in Swai et al. (2026).[43]

| | INTERFACE (asym) | CLAYFF (asym) | CLAYFF (sym) | GMINFF (asym) | GMINFF (sym) |
|---|---|---|---|---|---|
| PMF min. [kJ/mol] | -11.7 ± 1.9 | -20.7 ± 2.3 | -18.7 ± 1.3 | -21.4 ± 2.0 | -19.2 ± 2.2 |
| min. distance [nm] | 0.70 | 0.69 | 0.75 | 0.77 | 0.82 |
| $K_d$ [cm$^3$/g] | 16.4 | 326.0 | 161.8 | 360.2 | 138.9 |

All profiles show the two adsorption features, yet with different depths in the PMF profiles, with the global minimum located 0.69 to 0.82 nm from the layer centre, corresponding to 0.36 to 0.48 nm from the average position of the basal oxygen sites. CLAYFF and MINFF agree closely in both profile shape and well depth, although the MINFF minima lie more than 0.05 nm further from the surface than those obtained with CLAYFF or INTERFACE. The differences between CLAYFF and MINFF remain within the estimated uncertainties (95% confidence interval, n = 10), yet interestingly both force fields respond to the different ditrigonal cavities in the same way, since native asymmetric cavities (solid lines in Figure 7) place the adsorption minimum ca. 0.05 nm closer to the surface and deepen it by ca. 2 kJ/mol relative to the relaxed symmetric cavities (dashed lines in Figure 7). In comparison to experiments, this would correspond to a factor of more than 2 in the corresponding sorption coefficient $K_d$, obtained by integrating the profiles (see Swai et al., 2026 for details). Hence, in analogy with inorganic solutes,[19,95] the geometry of the ditrigonal cavities can affect the adsorption of organic solutes, and a force field that relaxes into symmetric cavities may underestimate the binding free energy by an amount comparable to thermal energy at ambient temperatures. Nevertheless, the differences in these PMF profiles across force fields and cavity geometries were found to be smaller than those between water models. In fact, additional simulations using MINFF (asym) with different water models (see Figure S8, Table S4) produced statistically similar PMF profiles between SPC/E, TIP3P, and OPC3 water models, whereas TIP3P/CHARMM produced a significantly shallower PMF profile. Among the three-site models, SPC yielded the deepest PMF profile, yet was shallower than TIP4P/2005 and OPC.

**7. Discussion**

Because the partial charges are shared between all parameter sets and all minerals, with only the Lennard-Jones parameters differing, MINFF with its charges derived across several representative and well-characterized minerals can be considered as a consistency test for Si, Al, Mg, Ca, Ti, Li, and F, for which the DDEC6 values scatter by less than 5% of their means. The nearest-neighbour charge-smearing rule of Eq. 6 assures this, since it derives every oxygen charge from the local bonding topology and thereby accommodates isomorphic substitution, edge sites, and protonation states automatically. $Fe^{2+}/Fe^{3+}$ is the exception that marks the limit of this approach, due to the wide range of DDEC6 charges for $Fe^{2+}/Fe^{3+}$ found for the different minerals having different coordination and substitution environments (Figure S1). Hence, the fixed +1.184 and +1.5 charges adopted in MINFF make up a simple compromise aimed at smectites and other silicates, where Fe is usually a minor component. The cost of this is visible in the underestimated bulk moduli of Hematite, Maghemite, and Wüstite, which is what a fixed-charge rigid-ion description of a partly covalent transition-metal–oxygen bond should be expected to give.

Overall, the general benchmark in Section 4 quantitatively shows the trade-off between general and tailored parameterisation. GMINFF, optimised synchronously across the training set of >30 reference minerals, produces mean lattice-parameter deviations of *ca.* 0.5% and median atom-type distance errors of 0.85–1.4%, whereas the TMINFF-M and -MO parameter sets, optimised mineral by mineral within the same functional form, reach *ca.* 0.13–0.49% and 0.07–0.37% in lattice-parameter deviations, and 0.72–1.82% and 0.72–1.58% in median atom-type distance errors, respectively. Hence the penalty for transferability amounts to a factor of ca. three or more in the a and b lattice metrics, but only ca. 1.1–1.5 along c, whereas the interatomic M–O distances remain comparatively insensitive, even when all Lennard-Jones parameters are tailored mineral by mineral. The same ordering appears in the diffraction benchmark, where for Kaolinite the GMINFF $R_{\mathrm{wp}}$ falls from 40.9% at $k_a$ = 0 to 18.8% at $k_a$ = 250 kJ·mol$^{-1}$·rad$^{-2}$, while the tailored TMINFF-M set reaches 9.4% at $k_a$ = 500, compared with 23.6% for CLAYFF and 42.6% for INTERFACE. However, tailored parameters are not uniformly beneficial but mineral-specific. The TMINFF-M set for Pyrophyllite at $k_a$ = 0 gave the lowest $G_{\mathrm{II}}$ of all Pyrophyllite runs together with the highest $R_{\mathrm{wp}}$ of the entire benchmark, 52.6%, since $\chi^2$ in Eq. 7 weights pair distances and angles differently to cell metrics and may favour local coordination over lattice parameters.

Because the oxygen Lennard-Jones parameters mediate most of the interaction between a mineral surface and water or an adsorbed molecule, their transferability was also examined separately. Interestingly, after the full TMINFF-MO optimisation they changed only slightly. For the O atom-types the tailored $\sigma$ varies across the training minerals with a coefficient of variation of 1.2–2.2% (median 1.3%) and stays within −5.0 to +3.7% of the OPC3 value, and $\epsilon$ within −4.2 to +3.0%, whereas the tailored metal parameters vary with a median coefficient of variation of 5.3% in $\sigma$ and 6.2% in $\epsilon$, and by up to 13.4% and 23.2% for individual atom types (Figure S9 and Table S5). The oxygen coordination environment is thus much more uniform across oxides and silicates than the cation environment, so that a single oxygen parameter pair is close to optimal everywhere, whereas a single metal pair is an average over structurally distinct sites. In practice, even the tailored MINFF sets present essentially the OPC3 oxygen to the aqueous phase, which turns the compatibility with OPC3-compatible water, ion, and organic force fields from an assumption made by construction into a measured property.

A breakdown of the M–O distance errors by atom type over the 32 training minerals shows that 24–42% of the mean-square error is a constant deviation per atom type, and that tailoring the Lennard-Jones parameters mineral by mineral does not remove it (Table S5). At $k_a$ = 500 kJ·mol$^{-1}$·rad$^{-2}$, for instance, the Sit–O distances are too short by 0.8% in GMINFF and by 1.1–1.2% in the tailored sets, whereas the Alo–O distances are too long by 0.8–1.2% in all three sets (Table S5). Since these deviations persist despite full optimisation of both $\sigma$ and $\epsilon$, they are unlikely to stem from the Lennard-Jones terms and instead point to the fixed partial charges and the single transferable angle term. The deviation share falls with increasing $k_a$, most clearly for TMINFF-M (from 39% at $k_a$ = 0 to 24% at 1500 kJ·mol$^{-1}$·rad$^{-2}$), hence part of the systematic error is taken up by the angle term rather than by the repulsive wall. The hydroxyl oxygen Oh, present in 17 of the training minerals, is the atom type for which the third optimisation step matters most: its RMS distance error is 2.68% in GMINFF and 2.74% in TMINFF-M, the largest of any atom type, and falls to 1.70% in TMINFF-MO, a reduction of ca. 38% relative to TMINFF-M and, at 1.04 percentage points, the largest absolute reduction of any atom type.

As expected, across both parameter sets the structural error decreases with increasing $k_a$, but not at the same rate for every property. For GMINFF the mean lattice deviations stay within 0.46–0.63% per axis between $k_a$ = 0 and 1500 kJ·mol$^{-1}$·rad$^{-2}$, whereas the Global Instability Index $G_{II}$, serving as a validation factor, falls from 0.228 to 0.111, and the median angle error falls from 0.27 to 0.11%. The angle strain analysis using the mean squared deviation $\langle(\theta-\theta_0)^2\rangle$ shows that increasing $k_a$ from 250 to 1500 kJ·mol$^{-1}$·rad$^{-2}$ reduces the strain by a factor of 2.5–6.8, with the smallest reductions found for the simple binary oxides, whose coordination geometry is already held close to ideal by close packing alone, and the largest for the open oxyhydroxide frameworks, which illustrates where in the training set the angle term does most of its work. The GMINFF metal Lennard-Jones parameters themselves depend only weakly on $k_a$: Sit $\sigma$ increases by 15.6% (and $\epsilon$ by 7.9%) from $k_a$ = 0 to 1500 kJ·mol$^{-1}$·rad$^{-2}$, whereas no other metal $\sigma$ changes by more than 3.2% (Table S2). A stiffer angle term around Sit thus takes over part of the work done by the Si–O repulsion at $k_a$ = 0, while for the other atom types the angle term and the Lennard-Jones terms are largely decoupled.

The diffraction tests are a useful complementary test to unit cell metrics and M–O bonding distances. Relative to CLAYFF, GMINFF gave equal or better agreement with the reference structures in 75% of the tested minerals, and relative to INTERFACE in 50%, while the tailored sets matched or surpassed both except for Corundum, Brucite, and Portlandite. The cases where CLAYFF retains an advantage, Kaolinite with its polymorphs Dickite and Nacrite, and Periclase, are the structures against which CLAYFF was itself optimised.

Interestingly, the bulk moduli of Quartz and Cristobalite are overestimated by every model tested, as expected when the tetrahedral-tilt compression mechanism responsible for the anomalously low experimental values cannot be represented. For the phyllosilicate elastic constants the anisotropy is more discriminating than the bulk modulus, with INTERFACE overestimating the in-plane $C_{11}$, $C_{22}$, and $C_{66}$ because of its rigid silicate bonds and angles while CLAYFF, having no bonded terms, underestimates the shear-derived $C_{44}$, $C_{55}$, and $C_{66}$, whereas MINFF with a moderate angle constant reproduces both the stiff in-plane and the soft shear constants, although using split angle terms for M-O-M and O-M-O triplets would be ideal.

In the librational and lattice region of the Kaolinite power spectra, $k_a$ = 250–500 kJ·mol$^{-1}$·rad$^{-2}$ qualitatively improves the band positions compared to the reported KGa-1 bands at 792, 753, 698, 539, 471, and 431 cm$^{-1}$, whereas using the CLAYFF-like $k_a$ = 0 leaves the region as the single broad 400–800 cm$^{-1}$ feature. On the opposite side $k_a$ = 1500 kJ·mol$^{-1}$·rad$^{-2}$ introduces blue-shifted bands of the kind seen with the highly strained INTERFACE case. However, significant differences were also found between the GMINFF and TMINFF parameter sets, even at the same $k_a$, which illustrates the sensitivity and potential use of the vibrational properties in further force field development.

In terms of clay mineral hydration, the Na-Montmorillonite case study reveals both differences and similarities between GMINFF and the CLAYFF and INTERFACE force fields. At the onset of smectite swelling at the lowest levels of hydration, CLAYFF gave very large differences in immersion energy between cases with hexagonal and trigonal basal-plane cavities, whereas GMINFF with angle terms using $k_a$ = 500 kJ·mol$^{-1}$·rad$^{-2}$ did not. Hence cavity geometry alone does not account for this. What matters instead is how readily the basal plane forms inner-sphere complexes with the interlayer cations, burying them deep within the basal-plane cavities. For the flexible, non-bonded CLAYFF description this burial is substantially deeper than for the angle-term restrained GMINFF, because the basal oxygens are less constrained and can therefore adapt to surface-adsorbed inner-sphere cations, as previously suggested by Underwood et al. (2016) and addressed by Ho et al. (2019).[19,95]

From the 2D lateral density maps of the 2W hydration state, it is found that MINFF preserves the native ditrigonal cavities (denoted asym) in the basal oxygen plane, and organises the surface water density into more distinct ridges over the basal plane cavities compared to the relaxed hexagonal cavities (denoted sym) in CLAYFF. Furthermore, converting the 2W density profiles into theoretical 00$l$ patterns ranks the models as GMINFF (asym) with OPC3 ($R_p$ = 0.005) ahead of CLAYFF (sym) with SPC/E (0.010) and well ahead of INTERFACE (asym) with SPC/E (0.069), indicating that the OPC3 oxygen parameters used throughout MINFF partially induce the same interlayer redistribution effect that the empirical CLAYFF-Mod basal-oxygen correction was introduced to achieve.[18]

The same effect carries over to the adsorption of organic molecules, as seen in the drug-molecule case, since CLAYFF and GMINFF respond to cavity geometry in the same direction and by a similar amount. The native ditrigonal cavities placed the Ciprofloxacin adsorption minimum *ca.* 0.05 nm closer to the surface and deepened it by *ca.* 2 kJ/mol, which corresponds to more than a factor of two in the integrated $K_d$. Set against this, the choice of water model is at least as consequential as the choice of mineral force field, since the global PMF minimum spans -11.5 to -28.2 kJ/mol across seven common water models (Table S4) against -19.2 to -21.4 kJ/mol for GMINFF between cavity geometries.

## 8. Conclusions and outlook

We have presented MINFF, a family of general classical force fields for clay minerals, oxides, and hydroxides, built on two design choices. The first is a single transferable harmonic angle term acting on all M–O (O–M–O and M–O–M) triplets, with different parameter sets at four levels of stiffness ($k_a$ = 0, 250, 500, and 1500 kJ·mol$^{-1}$·rad$^{-2}$), which controls coordination-polyhedron geometry and ditrigonal-cavity asymmetry while leaving the M–O distances to electrostatics and dispersion. The second is the extensive list of atom-types having newly derived DDEC6 cationic charges combined with the nearest-neighbour charge-smearing rule for O atom-types. This allows modelling of an arbitrary number of different defect sites, such as isomorphic substitution and unsaturated or saturated edge sites with different protonation states.

Across the benchmark library, the general parameter set reproduces unit-cell metrics to *ca.* 0.5–0.6%, atom-type distances to *ca.* 0.9–1.4%, and angles to *ca.* 0.1–0.3%, while lowering the Global Instability Index from 0.34 (CLAYFF) to 0.11–0.23, and below that of INTERFACE (0.21) for all non-zero angle constants, depending on $k_a$. Overall, GMINFF improves on CLAYFF for *ca.* two-thirds of the minerals tested and the tailored sets for more than 80% of them, and both match INTERFACE on the structures for which both are designed, while especially for the silicates offering improved vibrational properties. The presented case studies on Kaolinite and hydrated Na-Montmorillonite show that these numbers translate into the physical observables that motivated this further development of classical force fields, since with the additional angle terms the asymmetric ditrigonal cavity is well preserved rather than relaxed away, affecting the basal plane water organization, while producing basal spacings and immersion energies in line with expected values. For the adsorption of a zwitterionic antibiotic at the same surface, cavity geometry is worth ca. 2 kJ/mol in the free energy minimum, which translates to a factor of two in the $K_d$ sorption coefficient. However, the choice of water model, counterions, and other simulation parameters may lead to larger discrepancies.

Despite the broad efforts underpinning MINFF, several limitations remain and define the agenda for ongoing work. The benchmark is broad in chemistry but still weighted toward static structural observables and toward dynamical and energetic tests on smectites; systematic benchmarks of elastic constants, surface energies, and adsorption thermodynamics across the whole mineral library are in progress. The current release covers Si, Al, Fe, Mg, Ca, Ti, Li, and F-bearing minerals, and extension to further 3d transition metals such as Mn, Co, and Ni, to layered double hydroxides with carbonate or sulfate interlayers, and to phosphate and sulfate minerals is a natural next step that the present workflow should support without changes to the functional form. Furthermore, using a split between angle terms for O-M-O and M-O-M triplets may improve the bulk moduli, and the elastic and vibrational properties substantially. The treatment of $Fe^{2+}/Fe^{3+}$ remains the weakest part of the current parameterisation, since a single fixed charge per oxidation state cannot capture the range of the different charges found across different coordination and substitution environments, where an oxidation-state- and coordination-resolved assignment analogous to the Mgo/Mgh and Cao/Cah split is one future way to improve it. Finally, the parameters were fitted at ambient conditions against room-temperature or low-temperature refinements, so their transferability to high-pressure and high-temperature regimes has not been tested.

Overall, MINFF is intended as a pragmatic, general-purpose classical force field for clay minerals, oxides, and hydroxides. Its low computational cost, its interpretability, and its explicit handling of substitutions, edges, and framework topology make it suited to the time and length scales required for mineral-water interface, ion exchange reactions, and contaminant sorption, since its compatibility with OPC3 water and many corresponding ion-pair and organic potentials means that mineral, solvent, and solute can be treated within one consistent parameterisation. Its ability to reproduce experimental structures also makes it a usable source of labelled configurations for training and benchmarking machine-learning interatomic potentials on substitutionally disordered mineral systems, which remain difficult to generate from DFT alone at the required system sizes. MINFF could even potentially serve as the classical component of hybrid QM/MM or even ML/MM schemes, provided a machine-learned model trained on inorganic structure databases.[96,97] A fully non-bonded treatment of the metal sites would be an advantage in this context, since no M-O bond stretches need to be cut at the boundary, in contrast to force fields with explicit M–O bonds. All MINFF parameter files, reference structures, and topology tools are openly available via github.com/mholmboe/minff and the online topology server www.atomipy.io.

**Supporting Information**.

Supporting_information.pdf: Supporting information file including the General MINFF v1.0 parameter set, calculated partial charges for Fe using Spin-polarized DFT, calculated strain and relaxation factors, a meta-analysis of the Lennard-Jones parameter sets (atom-type distance error decomposition and spread of the tailored parameters), hydration properties of Na-Montmorillonite, comparison of *00l* 1D-XRD profiles for the 2W hydration state and umbrella sampling simulations with GMINFF vs. water models (PDF).

MINFF_v1.xls: The full parameter set of GMINFF, TMINFF-M and TMINFF-MO, including and benchmarking comparison with CLAYFF and INTERFACE force fields (XLS)

Short video showing how to generate a MINFF topology file using the online server at www.atomipy.io (MP4).

**Corresponding Author**

*Corresponding author: michael.holmboe@umu.se

**Data availability**

The supplementary information files contain all the general and tailored MINFF parameter sets and benchmarking data. Topology and simulation tools used to set up MINFF simulations are available from the first authors *atom*, *atomipy* and *minff* github repositories, respectively, as well as the online server *www.atomipy.io*. Any additional MD source data for this article are available upon request from the authors.

## Acknowledgement

This work was supported by the Swedish Research Council under grant number 2019-04733, and the computations were enabled by resources provided by the National Academic Infrastructure for Supercomputing in Sweden at HPC2N, Umeå University.

## TOC Graphical abstract

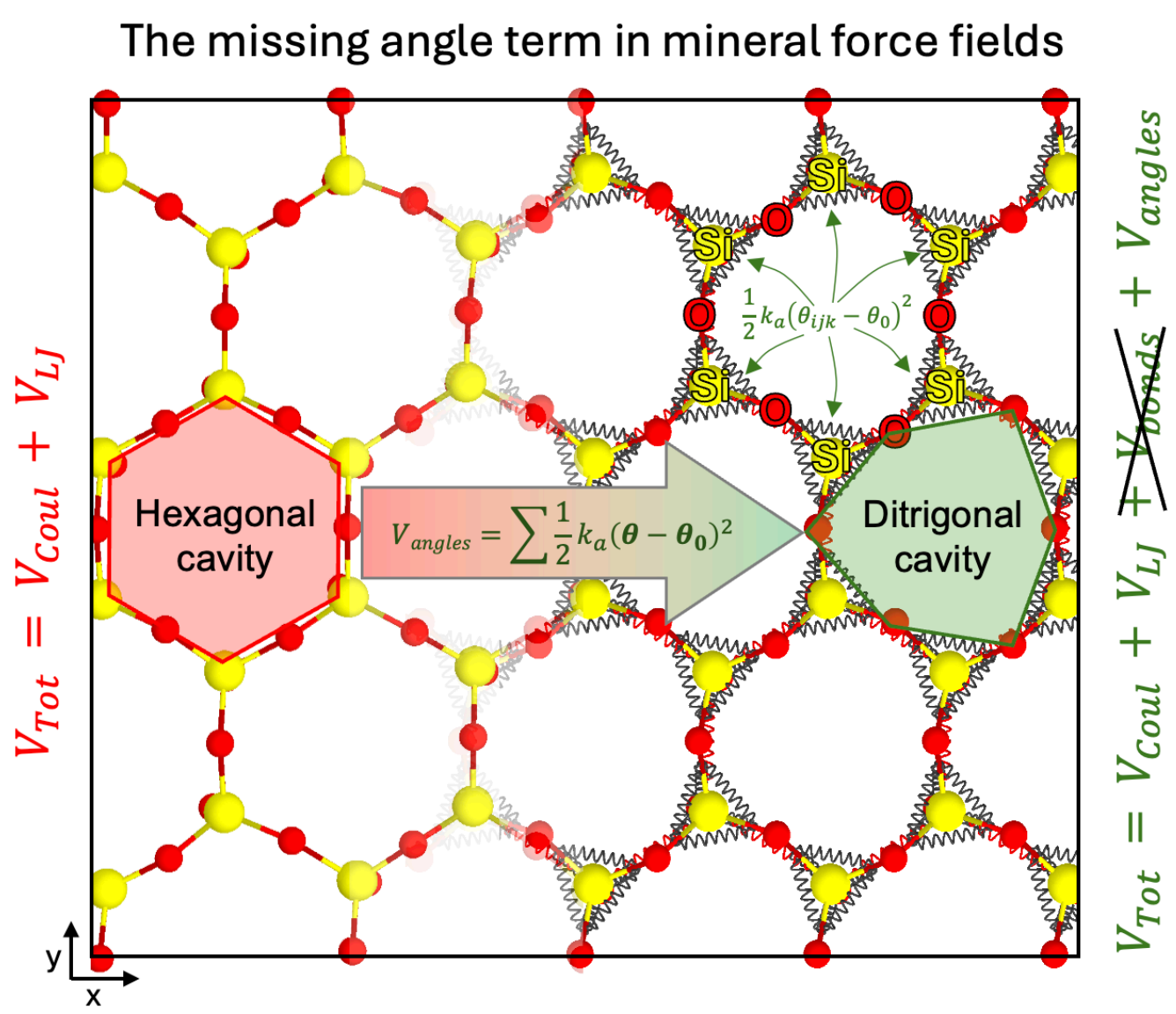

**MINFF: A Family of Classical Force fields for Minerals with Tunable Angle Terms**

Michael Holmboe*, Nathan Lavauzelle, Rogers Swai

*Department of Chemistry, Umeå University, SE-901 87 Umeå, Sweden*

*Corresponding author: michael.holmboe@umu.se

# Supporting information

## Table of contents

## 1. The General MINFF v1.0 parameter set

**Table S1**. The atom-types and M site charges of the MINFF v1.00 parameter set.

| type | mass [amu] | charge* | Comment |
|---|---|---|---|
| H | 1.00794 | 0.4 | Hydroxyl hydrogen |
| He | 1.00794 | 0.4 | Edge hydroxyl hydrogen |
| Lio | 6.94000 | 0.86 | Octahedral Lio, Ex. Li2O |
| Oahs | 15.99941 | -0.668 | Edge O, Al-OhH-Si |
| Oal | 15.99941 | -0.89517 | Edge O, Al-O |
| Oalh | 15.99941 | -1.197 | Edge O, Al-O-H |
| Oalhh | 15.99941 | -0.597 | Edge Al-OH2 oxygen |
| Oalsi | 15.99941 | -1.268 | Edge O, Al-O-Si |
| Oalt | 15.99941 | -1.2895 | Bridging (basal) O, Si-O-Al(t) |
| Ob | 15.99941 | -0.942 | Bridging oxygen, Ex. Si-O-Si |
| Obt | 15.99941 | -1.0865 | Bridging oxygen, Ex. Akdalaite |
| Odsub | 15.99941 | -1.325 | Bridging O, double iso. subst., Ex. Mg-O-Mg |
| Oh | 15.99941 | -0.994 | Hydroxyl oxygen |
| Ohmg | 15.99941 | -1.124 | Hydroxyl oxygen, iso. subst, Ex. H-O-Mg |
| Oli | 15.99941 | -0.89517 | Ex. Li2O |
| Omg | 15.99941 | -0.89517 | Bridging/apical O, Ex. in Al-O-Mg |
| Op | 15.99941 | -1.065 | Apical O, Ex. in Al-O-Si |
| Ops | 15.99941 | -0.942 | Apical O, Ex. in Al-O-Al(t) |
| Osi | 15.99941 | -0.89517 | Edge Si-O |
| Osih | 15.99941 | -0.871 | Edge Si-O-H |
| Fs | 18.99840 | -0.76/-0.87 | Ex. Fluorite (-0.76), hectorite (-0.87) |
| Mgh | 24.30500 | 1.74 | Ex. Brucite |
| Mgo | 24.30500 | 1.562 | Ex. Periclase |
| Mge | 24.30500 | 1.635 | Edge Mg2+, not rigorously tested |
| Alo | 26.98154 | 1.782 | Ex. Corundum, Montmorillonite |
| Ale | 26.98154 | 1.985 | Edge Al3+, not rigorously tested |
| Alt | 26.98154 | 1.782 | Tetrahedral Sit site in clays |
| Sit | 28.08538 | 1.884 | SiO2, clays |
| Cah | 40.07800 | 1.52 | Ex. Portlandite |
| Cao | 40.07800 | 1.66 | Ex. Lime |
| Fet2 | 55.84500 | 1.184 | Tetrahedral Fe2+, not rigorously tested |
| Fee2 | 55.84500 | 1.32 | Edge Fe2+, not rigorously tested |
| Feo2 | 55.84500 | 1.184 | Ex. Wustite |
| Fet3 | 55.84500 | 1.5 | Tetrahedral Fe3+ |
| Fee3 | 55.84500 | 1.75 | Edge Fe3+, not rigorously tested |
| Feo3 | 55.84500 | 1.5 | Ex. Hematite |
| Tio | 47.86700 | 2.48 | Ex. TiO2 |

Note: Oxygen charges must typically be calculated using Eq. 6.

**Table S2**. The atom-types and LJ-parameters of the general MINFF v1.00 parameter set.

| | ka=0 [kJ/mol/rad²] | | ka=250 [kJ/mol/rad²] | | ka=500 [kJ/mol/rad²] | | ka=1500 [kJ/mol/rad²] | |
|---|---|---|---|---|---|---|---|---|
| **Atom-type** | **sigma [nm]** | **epsilon [kJ/mol]** | **sigma [nm]** | **epsilon [kJ/mol]** | **sigma [nm]** | **epsilon [kJ/mol]** | **sigma [nm]** | **epsilon [kJ/mol]** |
| H | 0.000000 | 0.000000 | 0.000000 | 0.000000 | 0.000000 | 0.000000 | 0.000000 | 0.000000 |
| He | 0.000000 | 0.000000 | 0.000000 | 0.000000 | 0.000000 | 0.000000 | 0.000000 | 0.000000 |
| Lio | 0.191606 | 0.219589 | 0.192274 | 0.219667 | 0.191531 | 0.229417 | 0.192056 | 0.229375 |
| Oahs | 0.317427 | 0.683690 | 0.317427 | 0.683690 | 0.317427 | 0.683690 | 0.317427 | 0.683690 |
| Oal | 0.317427 | 0.683690 | 0.317427 | 0.683690 | 0.317427 | 0.683690 | 0.317427 | 0.683690 |
| Oalh | 0.317427 | 0.683690 | 0.317427 | 0.683690 | 0.317427 | 0.683690 | 0.317427 | 0.683690 |
| Oalhh | 0.317427 | 0.683690 | 0.317427 | 0.683690 | 0.317427 | 0.683690 | 0.317427 | 0.683690 |
| Oalsi | 0.317427 | 0.683690 | 0.317427 | 0.683690 | 0.317427 | 0.683690 | 0.317427 | 0.683690 |
| Oalt | 0.317427 | 0.683690 | 0.317427 | 0.683690 | 0.317427 | 0.683690 | 0.317427 | 0.683690 |
| Ob | 0.317427 | 0.683690 | 0.317427 | 0.683690 | 0.317427 | 0.683690 | 0.317427 | 0.683690 |
| Obt | 0.317427 | 0.683690 | 0.317427 | 0.683690 | 0.317427 | 0.683690 | 0.317427 | 0.683690 |
| Odsub | 0.317427 | 0.683690 | 0.317427 | 0.683690 | 0.317427 | 0.683690 | 0.317427 | 0.683690 |
| Oh | 0.317427 | 0.683690 | 0.317427 | 0.683690 | 0.317427 | 0.683690 | 0.317427 | 0.683690 |
| Ohmg | 0.317427 | 0.683690 | 0.317427 | 0.683690 | 0.317427 | 0.683690 | 0.317427 | 0.683690 |
| Oli | 0.317427 | 0.683690 | 0.317427 | 0.683690 | 0.317427 | 0.683690 | 0.317427 | 0.683690 |
| Omg | 0.317427 | 0.683690 | 0.317427 | 0.683690 | 0.317427 | 0.683690 | 0.317427 | 0.683690 |
| Op | 0.317427 | 0.683690 | 0.317427 | 0.683690 | 0.317427 | 0.683690 | 0.317427 | 0.683690 |
| Ops | 0.317427 | 0.683690 | 0.317427 | 0.683690 | 0.317427 | 0.683690 | 0.317427 | 0.683690 |
| Osi | 0.317427 | 0.683690 | 0.317427 | 0.683690 | 0.317427 | 0.683690 | 0.317427 | 0.683690 |
| Osih | 0.317427 | 0.683690 | 0.317427 | 0.683690 | 0.317427 | 0.683690 | 0.317427 | 0.683690 |
| Fs | 0.317622 | 0.192769 | 0.318354 | 0.189745 | 0.320267 | 0.185651 | 0.321464 | 0.190145 |
| Mgh | 0.188066 | 0.988654 | 0.186611 | 0.960819 | 0.186981 | 0.961886 | 0.190547 | 1.001225 |
| Mgo | 0.195549 | 0.734121 | 0.196144 | 0.744317 | 0.198028 | 0.716963 | 0.196465 | 0.742461 |
| Mge | 0.195549 | 0.734121 | 0.196144 | 0.744317 | 0.198028 | 0.716963 | 0.196465 | 0.742461 |
| Alo | 0.144095 | 0.473893 | 0.143862 | 0.466201 | 0.146158 | 0.471919 | 0.142145 | 0.460200 |
| Ale | 0.159028 | 0.369202 | 0.159028 | 0.369202 | 0.159028 | 0.369202 | 0.159028 | 0.369202 |
| Alt | 0.169243 | 0.178242 | 0.170069 | 0.176182 | 0.170704 | 0.175038 | 0.168338 | 0.175845 |
| Sit | 0.082225 | 0.422422 | 0.091168 | 0.443611 | 0.093467 | 0.458196 | 0.095058 | 0.455759 |
| Cah | 0.254786 | 0.872777 | 0.255577 | 0.908704 | 0.255225 | 0.889170 | 0.257601 | 0.909985 |
| Cao | 0.269310 | 1.577302 | 0.270104 | 1.511493 | 0.270147 | 1.577941 | 0.269989 | 1.509434 |
| Fet2 | 0.147279 | 0.171251 | 0.147279 | 0.171251 | 0.147279 | 0.171251 | 0.147279 | 0.171251 |
| Fee2 | 0.215896 | 0.230825 | 0.220018 | 0.210001 | 0.220434 | 0.210576 | 0.220810 | 0.210784 |
| Feo2 | 0.215896 | 0.230825 | 0.220018 | 0.210001 | 0.220434 | 0.210576 | 0.220810 | 0.210784 |
| Fet3 | 0.197627 | 0.119088 | 0.187131 | 0.113822 | 0.188425 | 0.115038 | 0.191267 | 0.116022 |
| Fee3 | 0.168158 | 0.385264 | 0.169854 | 0.370788 | 0.169565 | 0.387325 | 0.170260 | 0.381470 |
| Feo3 | 0.168158 | 0.385264 | 0.169854 | 0.370788 | 0.169565 | 0.387325 | 0.170260 | 0.381470 |
| Tio | 0.188206 | 0.806177 | 0.189407 | 0.806691 | 0.188503 | 0.808327 | 0.186107 | 0.794466 |

Note: Parameters for TMINFF-M and TMINFF-MO can be found at github.com/mholmboe/minff

## 2. Calculation of partial charges for Fe using Spin-polarized DFT

Spin-polarized electronic structure calculations were performed within the framework of Density Functional Theory (DFT), employing the Gaussian Plane Wave (GPW) method as implemented in the QUICKSTEP module of the CP2K code (VandeVondele et al. 2005; Kühne et al. 2020). To avoid the explicit treatment of core electrons, for iron corresponding to the [Ne] $3s^2$ $3p^6$ configuration, scalar-relativistic norm-conserving pseudopotentials of the Goedecker–Teter–Hutter (GTH) type were applied. The valence electron wave functions were expanded in a linear combination of contracted Gaussian-type orbitals using the MOLOPT basis sets, which are specifically optimized for use with GTH pseudopotentials (VandeVondele and Hutter 2007).

An electronic density cutoff of 1500 Ry was adopted for the density-of-states calculations, thereby determining the resolution of the auxiliary plane-wave basis set. Exchange-correlation energies were evaluated using the Perdew–Burke–Ernzerhof (PBE) functional. Spin multiplicities (2S + 1) of 5 and 6 were assigned to $Fe^{2+}$ and $Fe^{3+}$ ions, respectively, while a multiplicity of 2 was used for $Fe^{2+}$–$Fe^{3+}$ pairs. Since conventional DFT is known to underestimate the Coulomb repulsion among the localized 3d electrons of iron, the DFT+U correction was applied with a Hubbard parameter of U = 1.9 eV, as refined by Kéri et al. (Kéri et al. 2017).

The partial charges for $Fe^{2+}$ and $Fe^{3+}$ sites were selected empirically, with a primary focus on Fe-bearing smectites. Figure S1 presents the partial charges across a range of substitution environments, oxidation states, and coordination numbers for Fe-bearing smectites and reference oxy-hydroxides. A general trend emerges for $Fe^{3+}$, which consistently exhibits partial charges above +1.4, with the notable exception of Goethite, which displays a lower value and may require further charge optimization analogous to the approach described for Mg and Ca. In contrast, the partial charges for $Fe^{2+}$ are more broadly distributed, ranging from +0.8 to +1.4, with Wüstite yielding the highest value within this range.

Given that GMINFF was initially developed for clay minerals, and given that naturally occurring clay minerals typically contain only minor amounts of iron most commonly as isolated Fe atoms or small clusters, a practical and justified charge of +1.5 for $Fe^{3+}$ and +1.184 for $Fe^{2+}$ was adopted. This choice is further supported by the partial charge of +1.2 obtained for oxidised Nontronites, which is consistent with the value of +1.2 derived from $Fe^{2+}$–$Fe^{3+}$ pair calculations in smectites, indicating that partial charges from single Fe and Fe pairs are sufficient to describe various Fe contents in smectites.

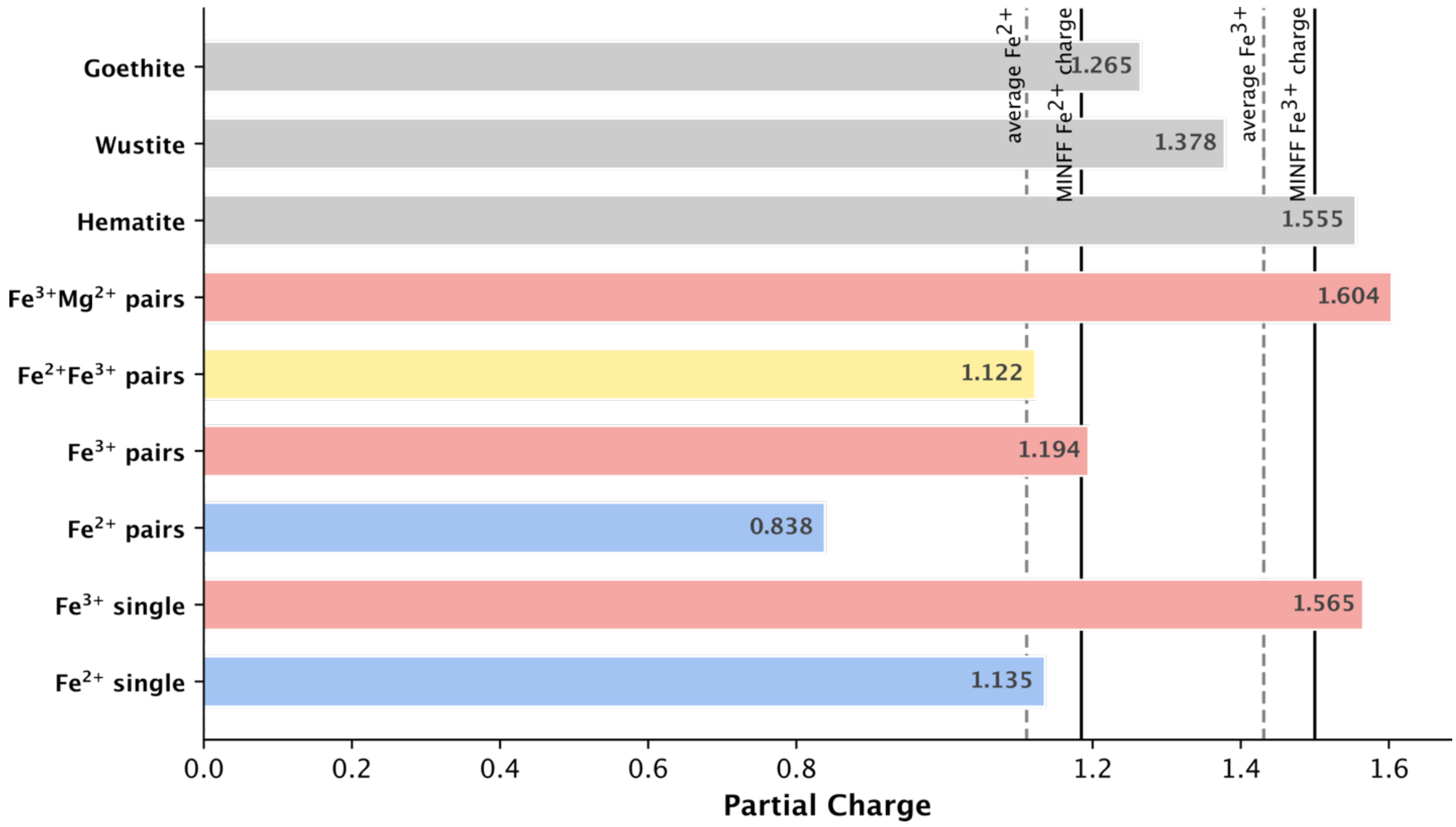

**Figure S1**. Partial charges calculated using DDEC6 on GEO-OPT DFT simulations for different Fe-bearing minerals, including Montmorillonite with $Fe^{2+}$ (blue), $Fe^{3+}$ (red), and $Fe^{2+}$-$Fe^{3+}$ mixed valence (yellow).

## 3. Calculation of strain and relaxation factors

**Table S3**. Angle strain factor from $\langle(\theta-\theta_0)^2\rangle/\left(\frac{1}{2}k_a N_{angles}\right)$ in units of rad$^2$ for the different tested minerals, versus $k_a$ in unit kJ·mol$^{-1}$·rad$^{-2}$. StrainRatio factor indicates the strain ratio of $k_{a-1500}/k_{a-250}$.

| Mineral | Class | ka=250 | ka=500 | ka=1500 | StrainRatio |
|---|---|---|---|---|---|
| CaO | oxide | 0.00069 | 0.00048 | 0.00022 | 3.153 |
| Periclase | oxide | 0.00070 | 0.00048 | 0.00022 | 3.122 |
| Rutile | oxide | 0.00087 | 0.00064 | 0.00034 | 2.541 |
| $Li_2O$ | oxide | 0.00089 | 0.00058 | 0.00026 | 3.498 |
| Wüstite | oxide | 0.00091 | 0.00058 | 0.00024 | 3.839 |
| Anatase | oxide | 0.00103 | 0.00073 | 0.00036 | 2.815 |
| $CaF_2$ | halide | 0.00104 | 0.00065 | 0.00027 | 3.837 |
| Corundum | oxide | 0.00110 | 0.00076 | 0.00038 | 2.867 |
| Forsterite | nesosilicate | 0.00124 | 0.00087 | 0.00041 | 3.042 |
| Hematite | oxide | 0.00130 | 0.00082 | 0.00035 | 3.694 |
| Brucite | hydroxide | 0.00132 | 0.00091 | 0.00044 | 2.998 |
| Talc | phyllosilicate | 0.00161 | 0.00116 | 0.00055 | 2.947 |
| Imogolite | phyllosilicate | 0.00215 | 0.00143 | 0.00065 | 3.284 |
| Portlandite | hydroxide | 0.00251 | 0.00164 | 0.00070 | 3.566 |
| Hectorite-H | phyllosilicate | 0.00254 | 0.00171 | 0.00072 | 3.548 |
| Akdalaite | oxyhydroxide | 0.00260 | 0.00147 | 0.00054 | 4.812 |
| Nacrite | phyllosilicate | 0.00276 | 0.00193 | 0.00084 | 3.282 |
| Dickite | phyllosilicate | 0.00289 | 0.00205 | 0.00089 | 3.249 |
| Gibbsite | hydroxide | 0.00311 | 0.00195 | 0.00076 | 4.103 |
| Hectorite-F | phyllosilicate | 0.00364 | 0.00225 | 0.00084 | 4.343 |
| Kaolinite | phyllosilicate | 0.00365 | 0.00245 | 0.00102 | 3.566 |
| Pyrophyllite | phyllosilicate | 0.00371 | 0.00244 | 0.00102 | 3.637 |
| 16-trans-Oct-Mg2cis-Fe3cis | phyllosilicate | 0.00380 | 0.00236 | 0.00091 | 4.174 |
| 17-trans-Tet-Fe3 | phyllosilicate | 0.00381 | 0.00239 | 0.00093 | 4.115 |
| 15-trans-Oct-Fe2-cis | phyllosilicate | 0.00384 | 0.00238 | 0.00092 | 4.188 |
| 9-cis-Oct-Fe2-trans | phyllosilicate | 0.00406 | 0.00250 | 0.00096 | 4.235 |
| 12-cis-Oct-Mg2trans-Fe3cis | phyllosilicate | 0.00406 | 0.00249 | 0.00096 | 4.240 |
| 11-cis-Oct-Mg2cis-Fe3trans | phyllosilicate | 0.00406 | 0.00251 | 0.00096 | 4.221 |
| 13-cis-Oct-Mg2trans-Fe3trans | phyllosilicate | 0.00409 | 0.00251 | 0.00096 | 4.252 |
| 14-cis-Tet-Fe3 | phyllosilicate | 0.00411 | 0.00255 | 0.00098 | 4.201 |
| Montmorillonite | phyllosilicate | 0.00411 | 0.00273 | 0.00113 | 3.631 |
| 8-cis-Oct-Fe2-cis | phyllosilicate | 0.00416 | 0.00256 | 0.00098 | 4.238 |
| Coesite | tectosilicate | 0.00420 | 0.00266 | 0.00107 | 3.943 |
| 10-cis-Oct-Mg2cis-Fe3cis | phyllosilicate | 0.00422 | 0.00257 | 0.00098 | 4.327 |
| Quartz | tectosilicate | 0.00427 | 0.00235 | 0.00092 | 4.653 |
| Boehmite | oxyhydroxide | 0.00433 | 0.00247 | 0.00072 | 6.017 |
| Cristobalite | tectosilicate | 0.00450 | 0.00232 | 0.00088 | 5.088 |
| Goethite | oxyhydroxide | 0.00460 | 0.00249 | 0.00084 | 5.505 |
| Lepidocrocite | oxyhydroxide | 0.00494 | 0.00239 | 0.00072 | 6.830 |
| Muscovite | phyllosilicate | 0.00516 | 0.00327 | 0.00130 | 3.981 |
| Diaspore | oxyhydroxide | 0.00533 | 0.00275 | 0.00092 | 5.770 |
| Nontronite | phyllosilicate | 0.00631 | 0.00390 | 0.00142 | 4.436 |
| Maghemite | spinel | 0.00943 | 0.00522 | 0.00159 | 5.937 |

## 4. Hydration properties of Na-Montmorillonite

Sequenced simulations as a function of diminishing water content was performed according to Figure S2, using Na-Montmorillonite having the formula $Na_{0.66}[Al_{3.34}Mg_{0.66}][Si_8]O_{20}(OH)_4)$, modelled as a 6x4x3 supercell with 4320 SPC/E water and ion-pair potentials from Joung and Cheatham 2008, as well as OPC3 with ion-pair potentials from Li and Merz (2021). For CLAYFF and GMINFF, the Montmorillonite structure was based on Lee and Guggenheim (1981), whereas for INTERFACE, the structure reported by Rothbauer (1971) was used.

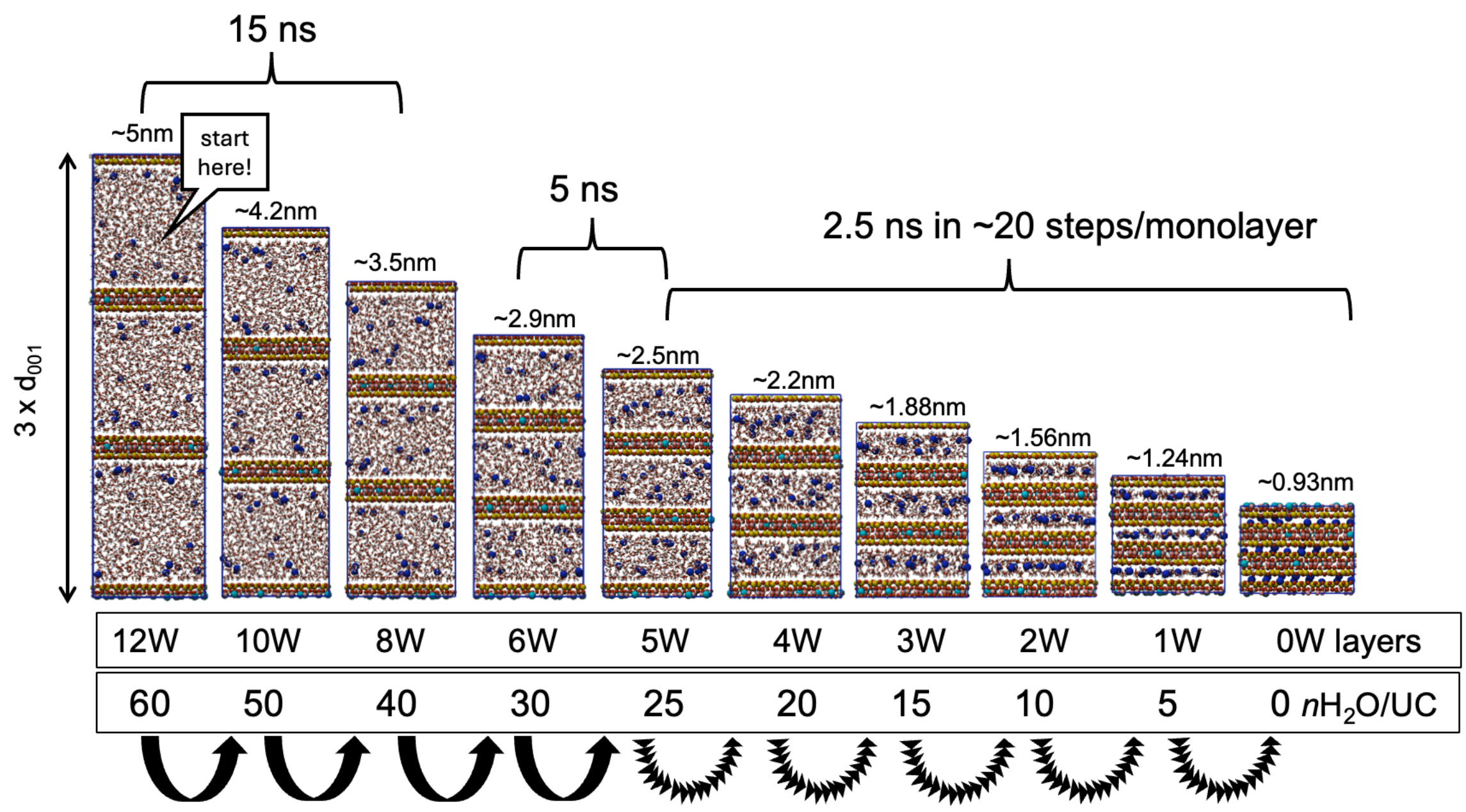

**Figure S2**. Schematic illustration of systems of Na-Montmorillonite sequentially simulated with decreasing water content after removal of fixed number of water molecules in each interlayer.

*4.1 Self-diffusion coefficients of water as a function of $d_{001}$*

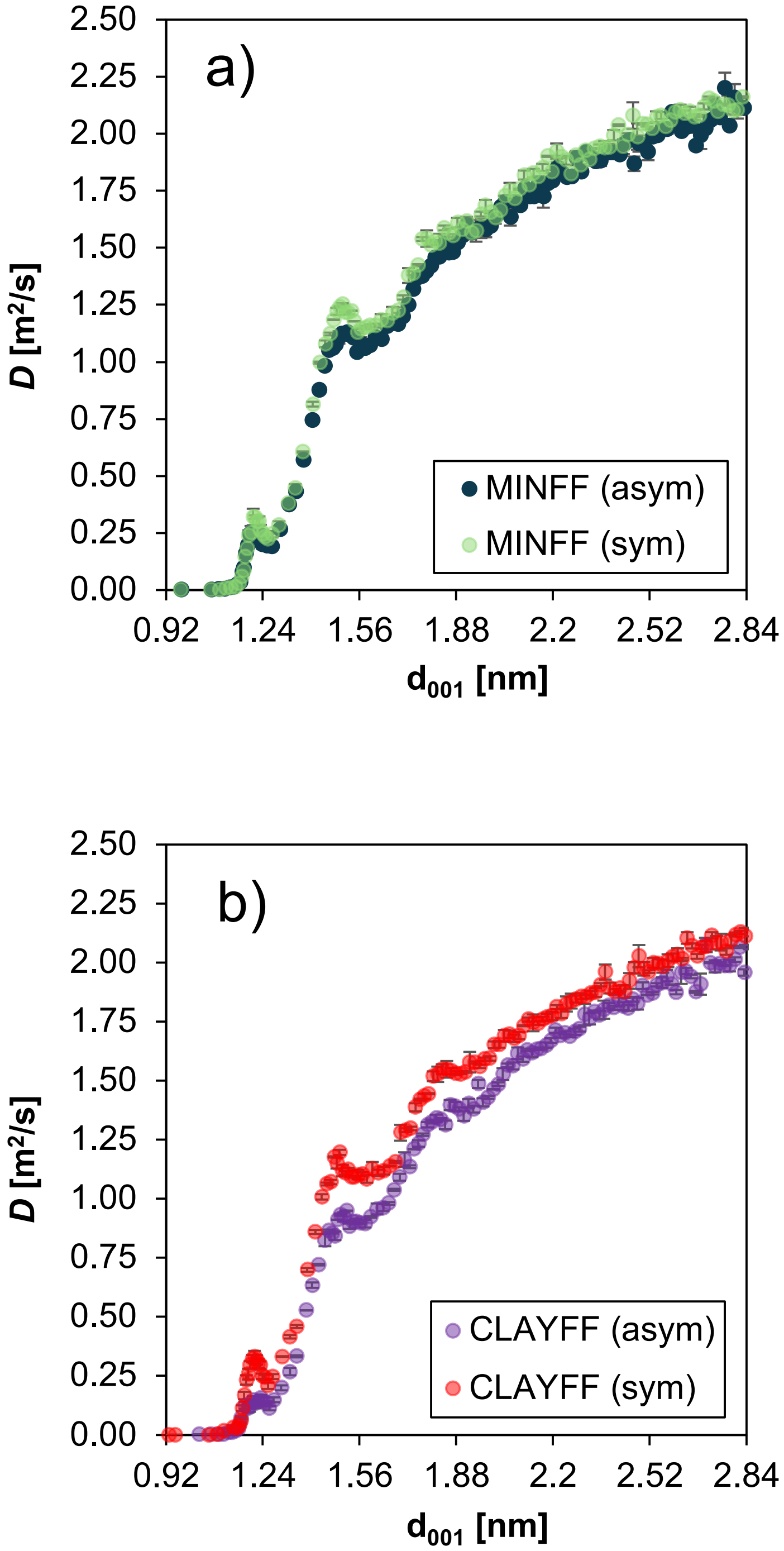

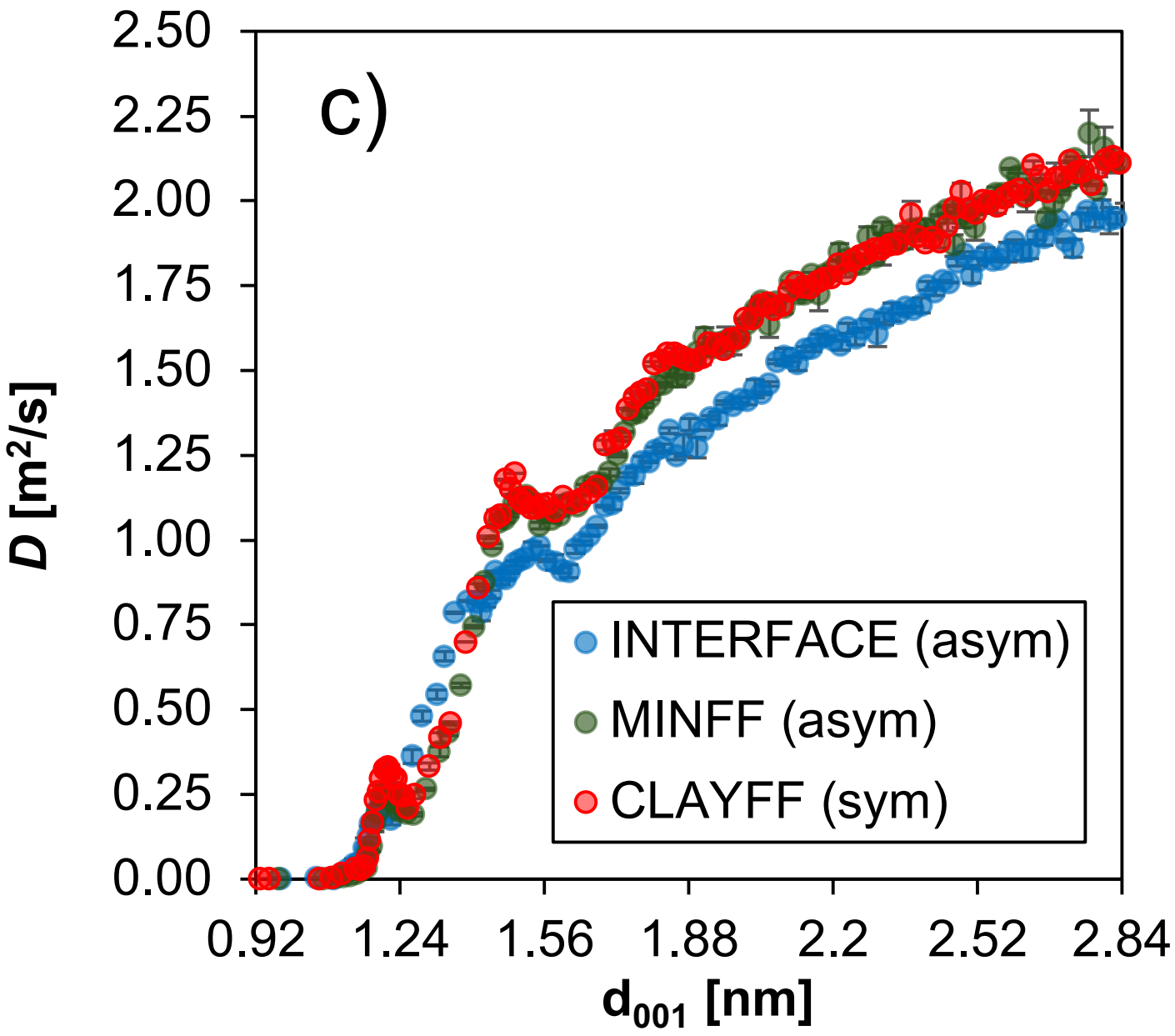

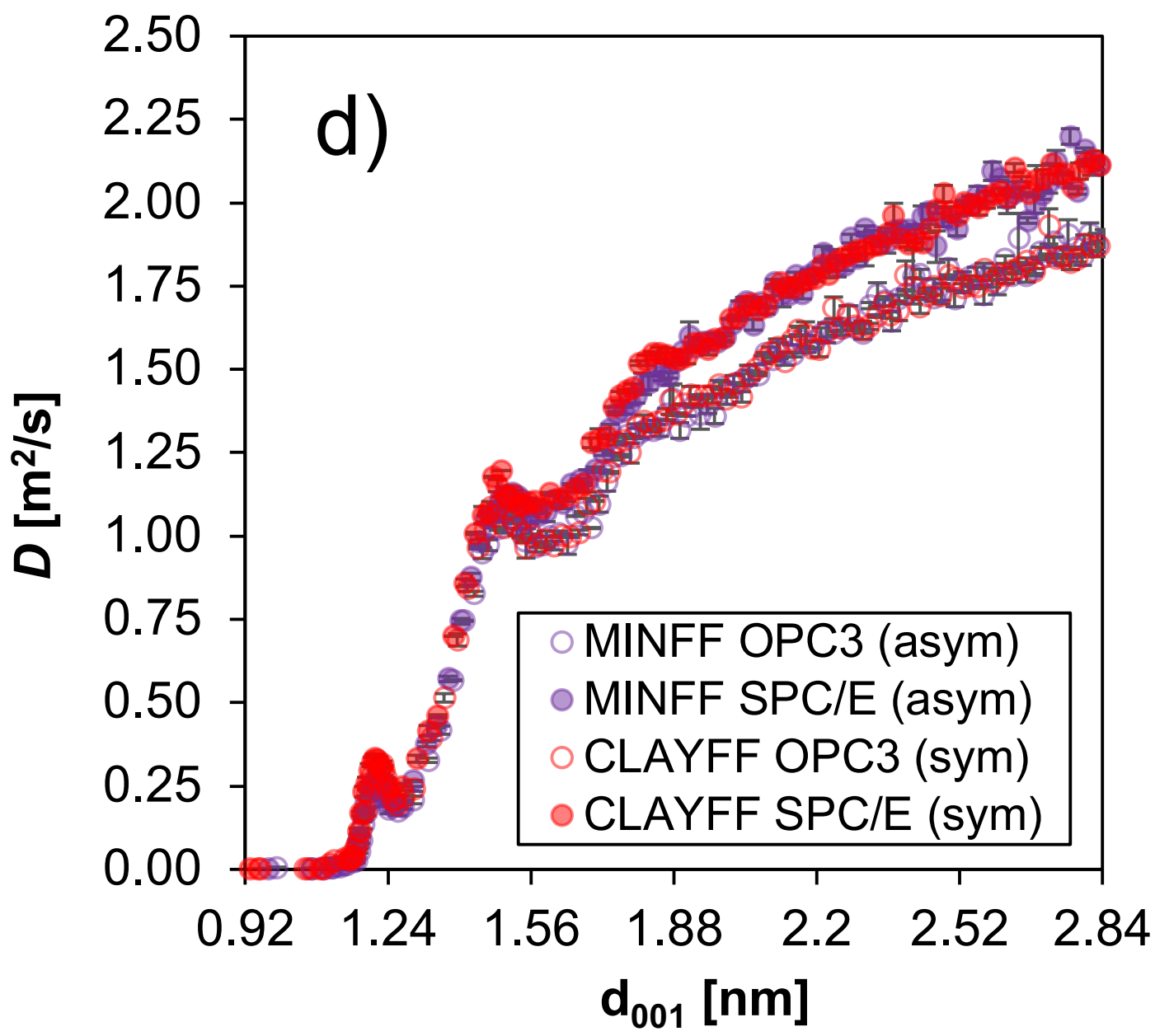

**Figure S3**. Self-diffusion coefficients of interlayer water in Na-Montmorillonite as a function of basal spacing $d_{001}$. a) GMINFF with hexagonal (sym) and ditrigonal (asym) cavities, set by $k_a$ = 500 kJ·mol$^{-1}$·rad$^{-2}$. b) CLAYFF with hexagonal (sym) and ditrigonal (asym) cavities. c) Comparison of GMINFF, CLAYFF and INTERFACE with ditrigonal cavities. d) Comparison between the SPC/E and newer OPC3 water model for CLAYFF (sym) and GMINFF (asym.)

### *4.2 Density maps of water in the 2W hydration state*

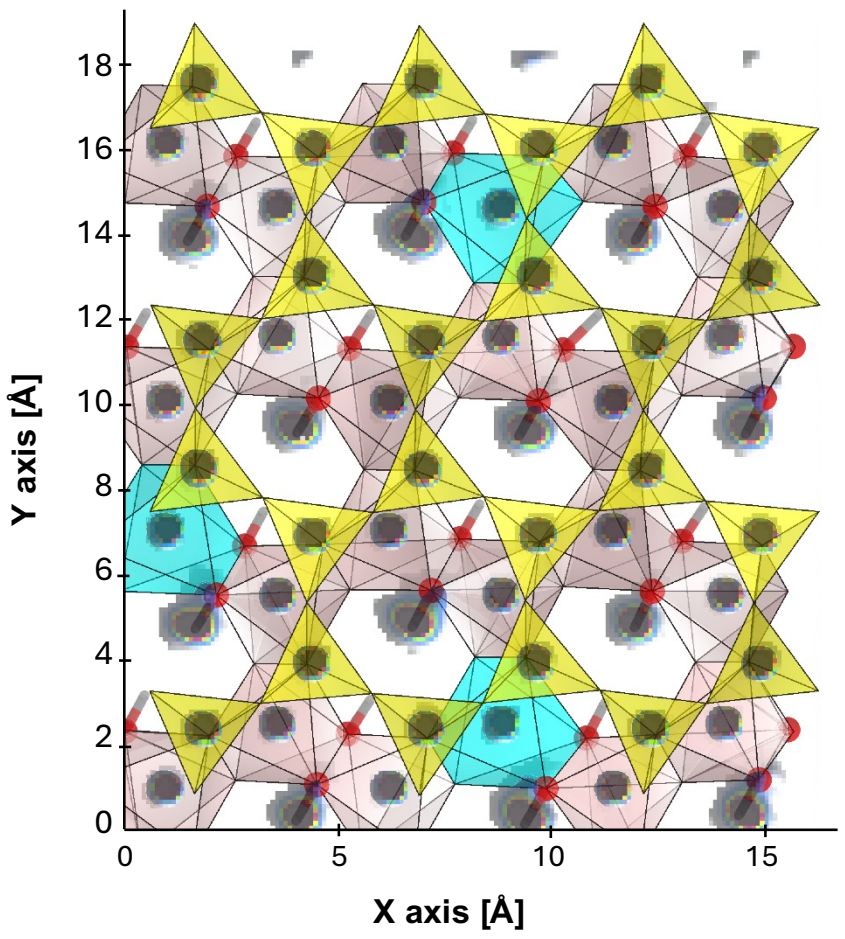

**Figure S4**. Top half structure of the modelled trans-vacant Na-Montmorillonite displaying the density sites of the tetrahedral Si (yellow tetrahedra), octahedral Al/Mg (pink/turquoise octahedra) and lattice hydroxyl groups. Note that the lower tetrahedral sheet of the TOT smectite structure is omitted for clarity.

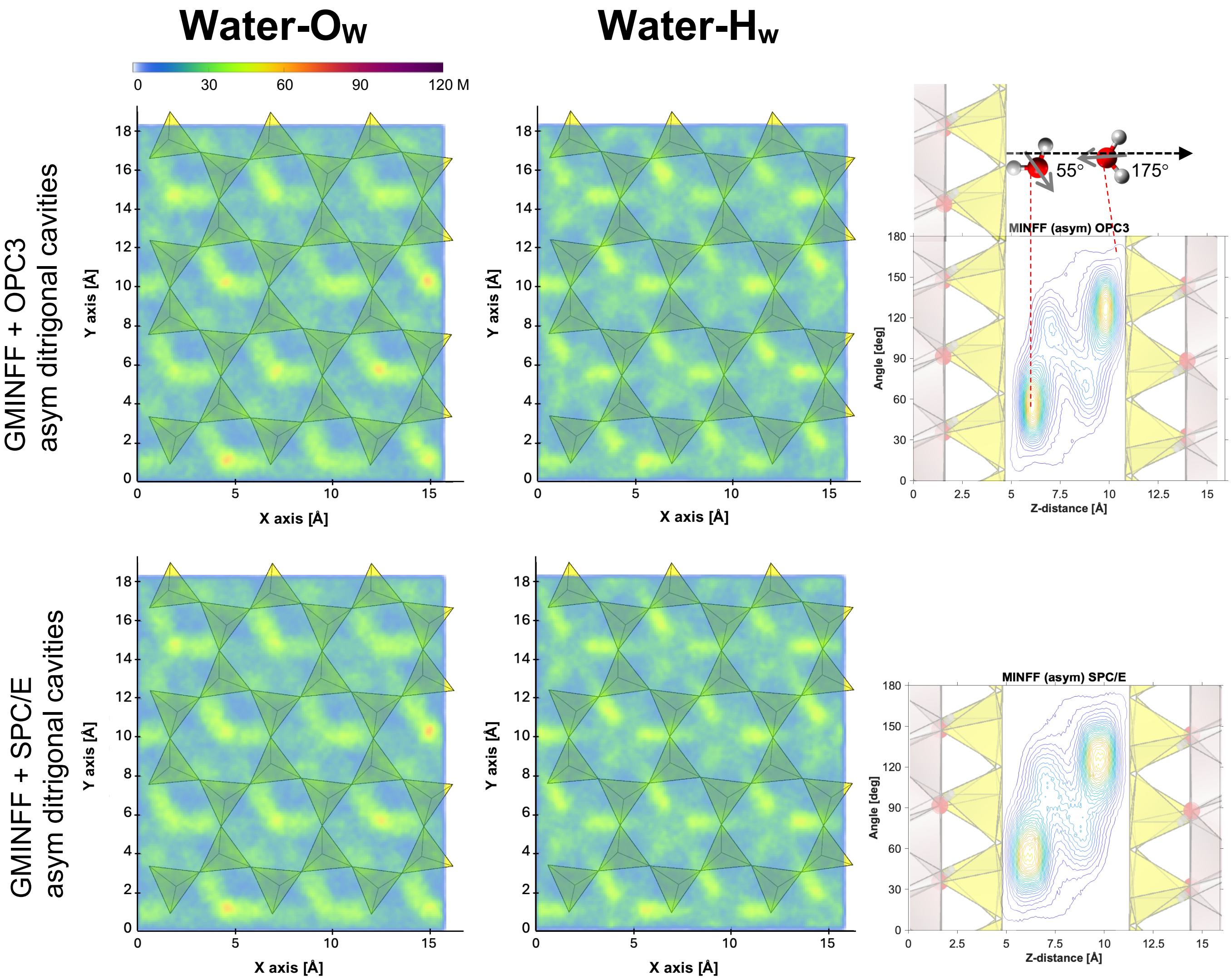

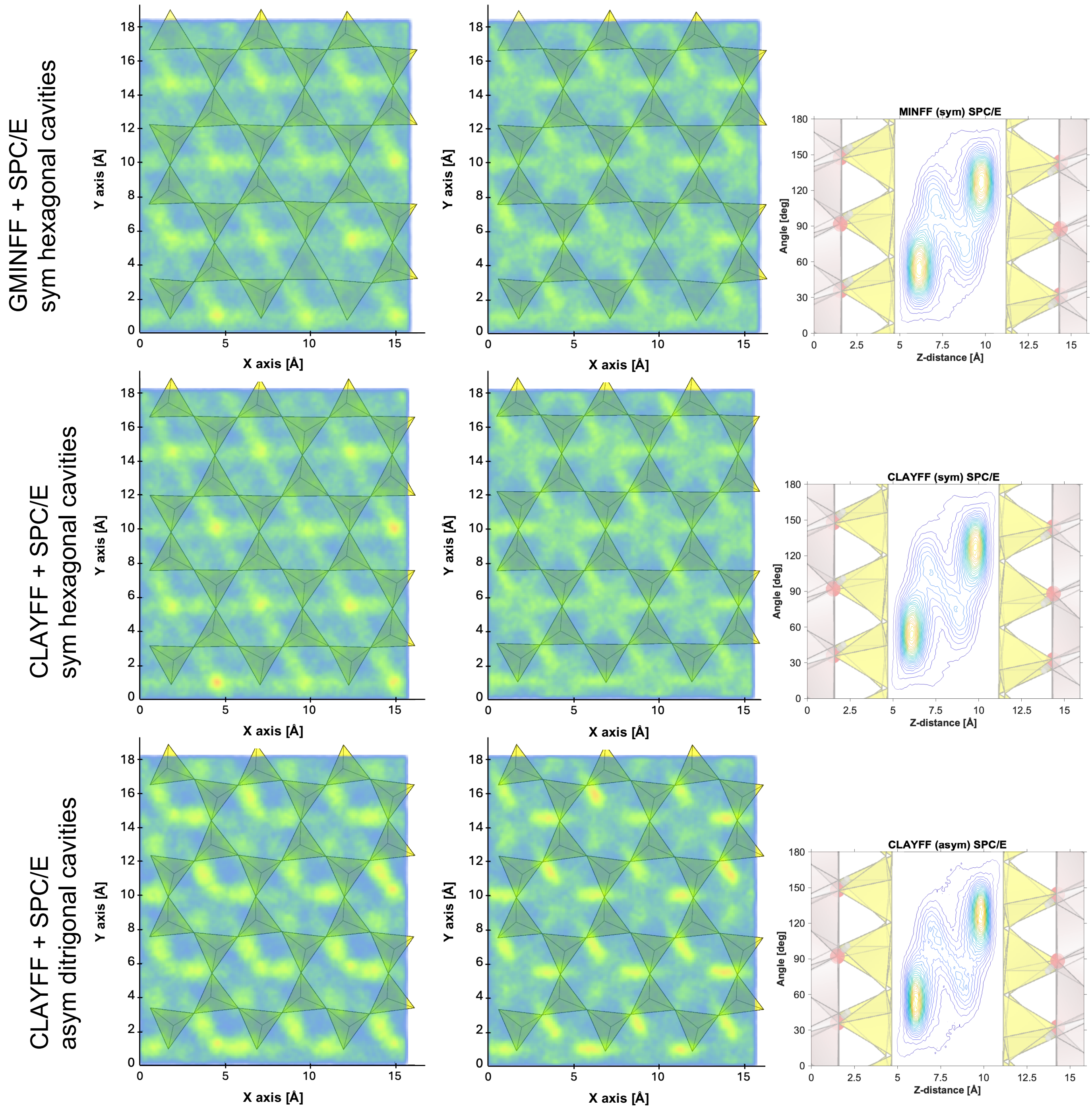
GMINFF + SPC/E
sym hexagonal cavities
CLAYFF + SPC/E
sym hexagonal cavities
CLAYFF + SPC/E
asym ditrigonal cavities
MINFF (sym) SPC/E
CLAYFF (sym) SPC/E
CLAYFF (asym) SPC/E
X axis [Å]
Y axis [Å]
Z-distance [Å]
Angle [deg]

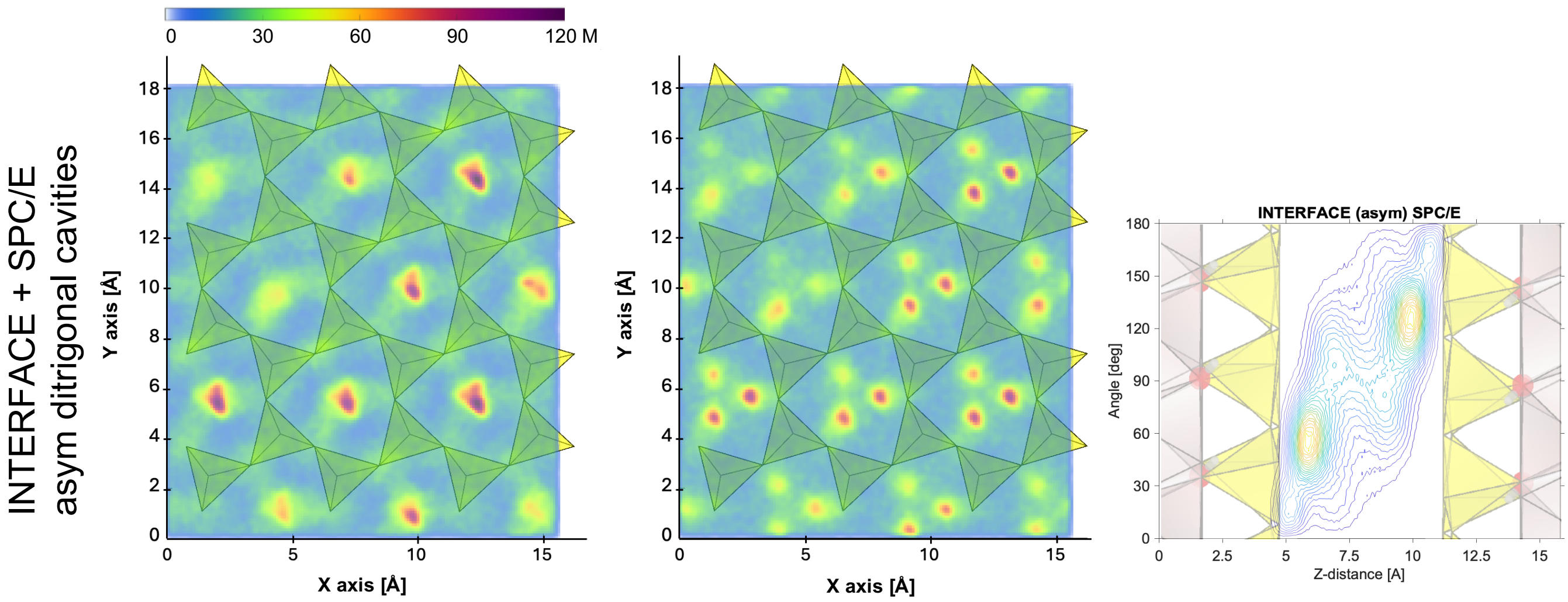

**Figure S5. Density maps** of water from half the 2W interlayer interfacing the Montmorillonite basal surface modelled with the general GMINFF and $k_a$ = 500 kJ·mol$^{-1}$·rad$^{-2}$. The left column shows the density of oxygen (Ow), and the middle column shows the density of water hydrogen (Hw), in molar concentrations. Right column show angular density maps of the interlayer water, defined as the angle to the left surface normal as indicated in the top/right image. Note that the top row shows GMINFF with OPC3 water; the rest are SPC/E. MINFF and CLAYFF structure based on Lee and Guggenheim (1981). INTERFACE structure based on Rothbauer, 1971.

*4.3 Density profiles of the 2W hydration state in Na-Montmorillonite*

Comparison of overlapping density profiles between pairs of 15.60 Å 2W Na-Montmorillonite systems modelled with GMINFF, CLAYFF, and INTERFACE, having different cavity geometries and SPC/E or OPC3 water models. See Figure S6 caption below for more details.

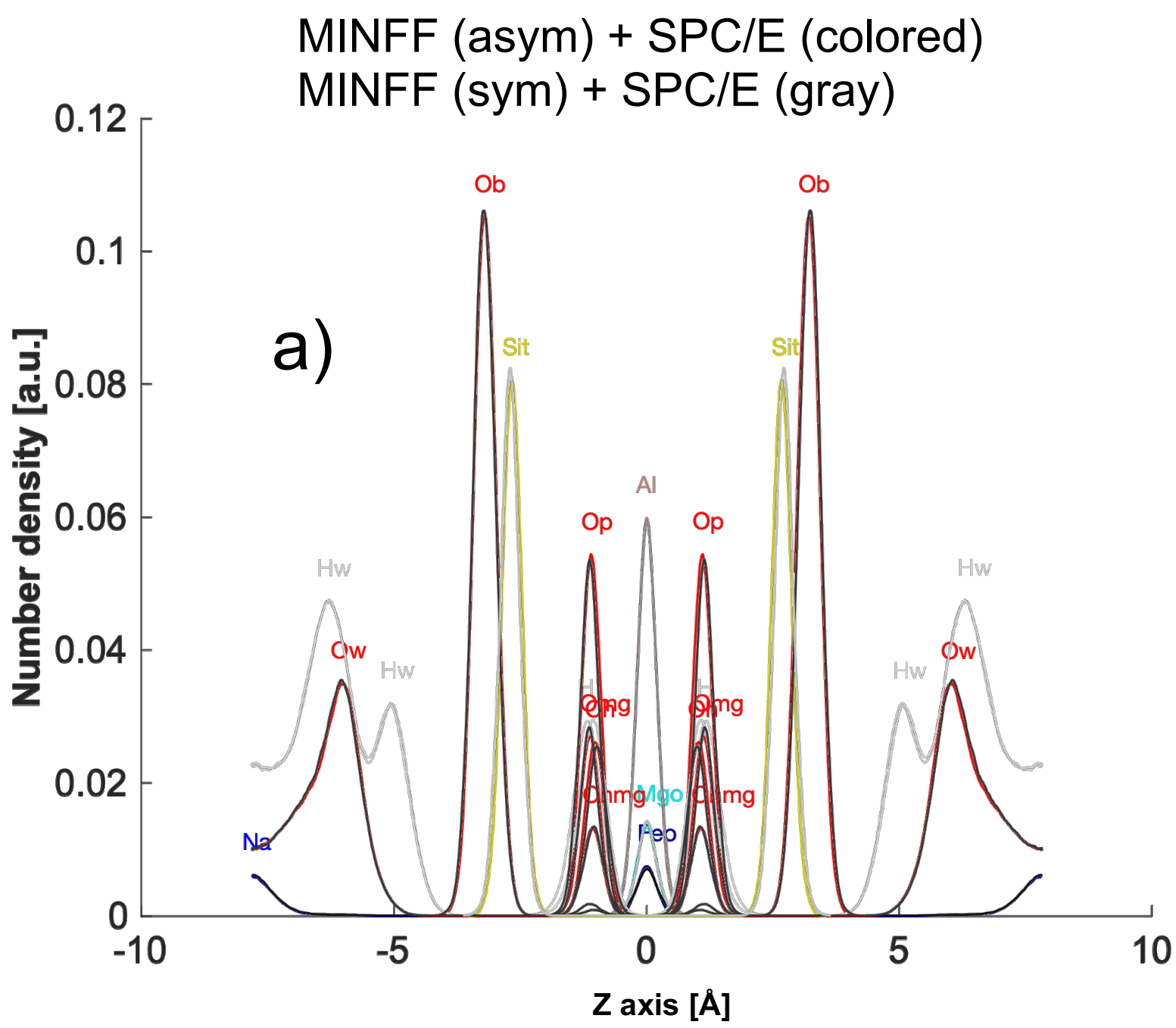

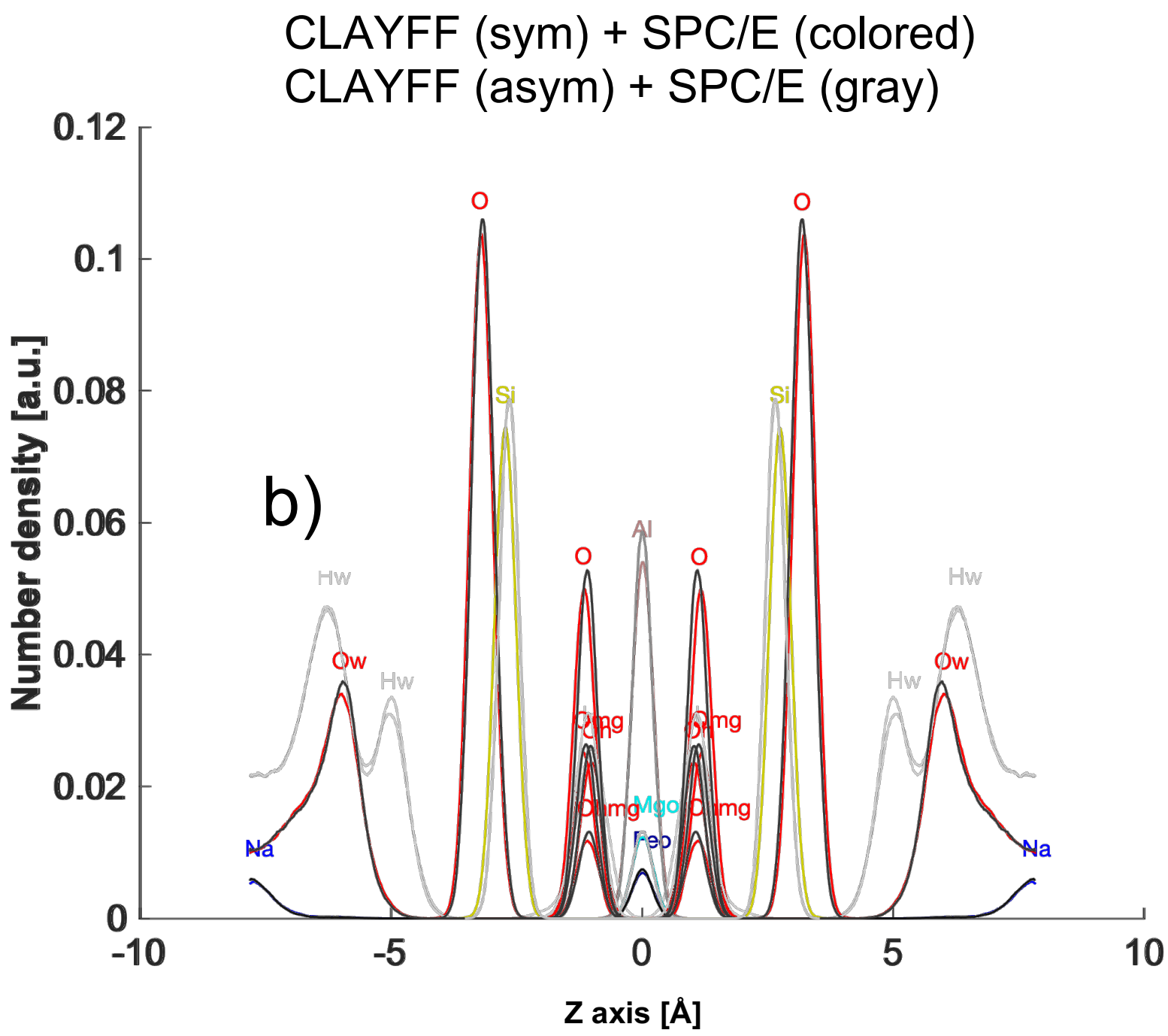

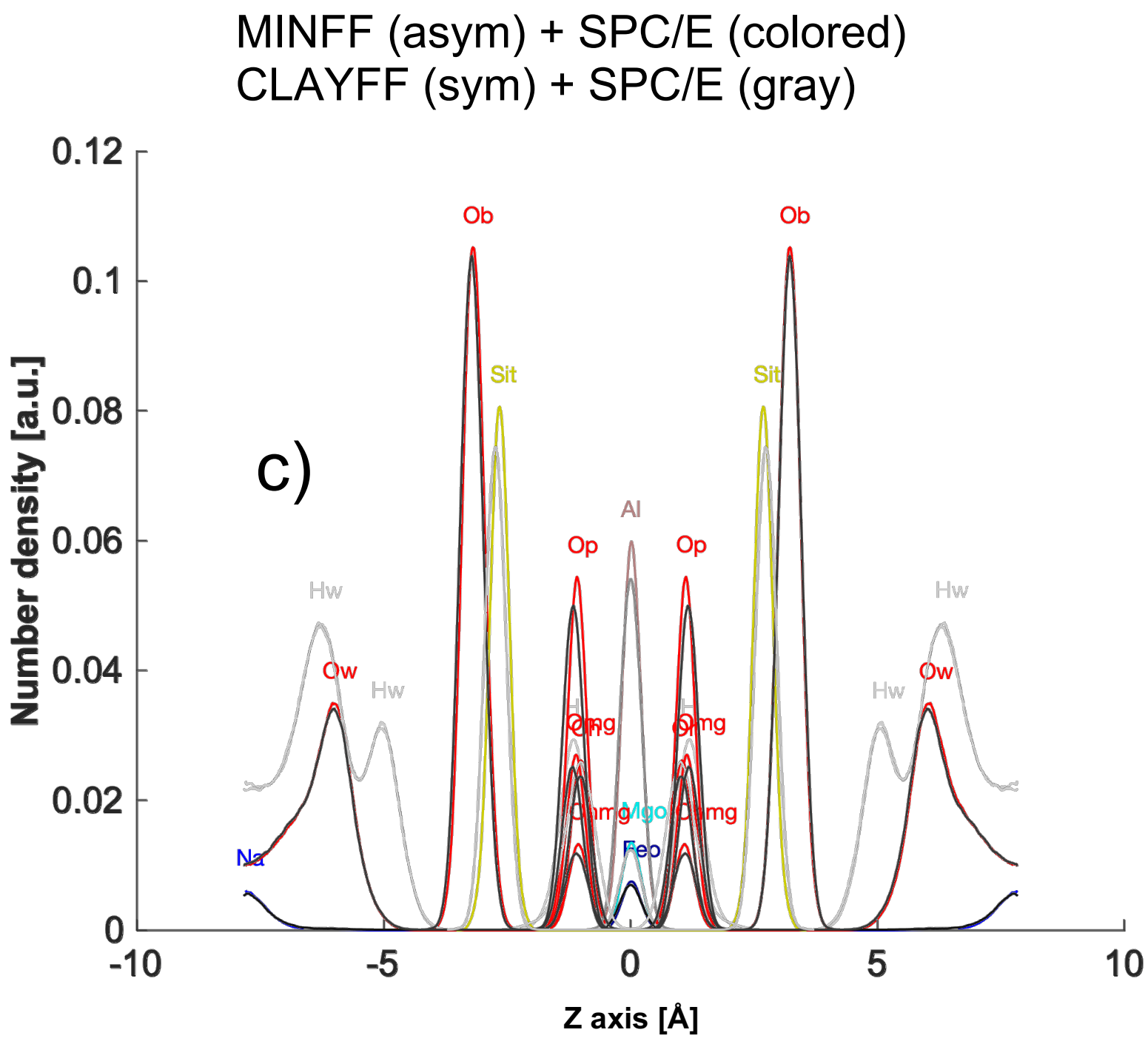
MINFF (asym) + SPC/E (colored)
CLAYFF (sym) + SPC/E (gray)
c)
Number density [a.u.]
Z axis [Å]
Ob
Sit
Op
Al
Hw
Ow
Na
Feo
Mgo

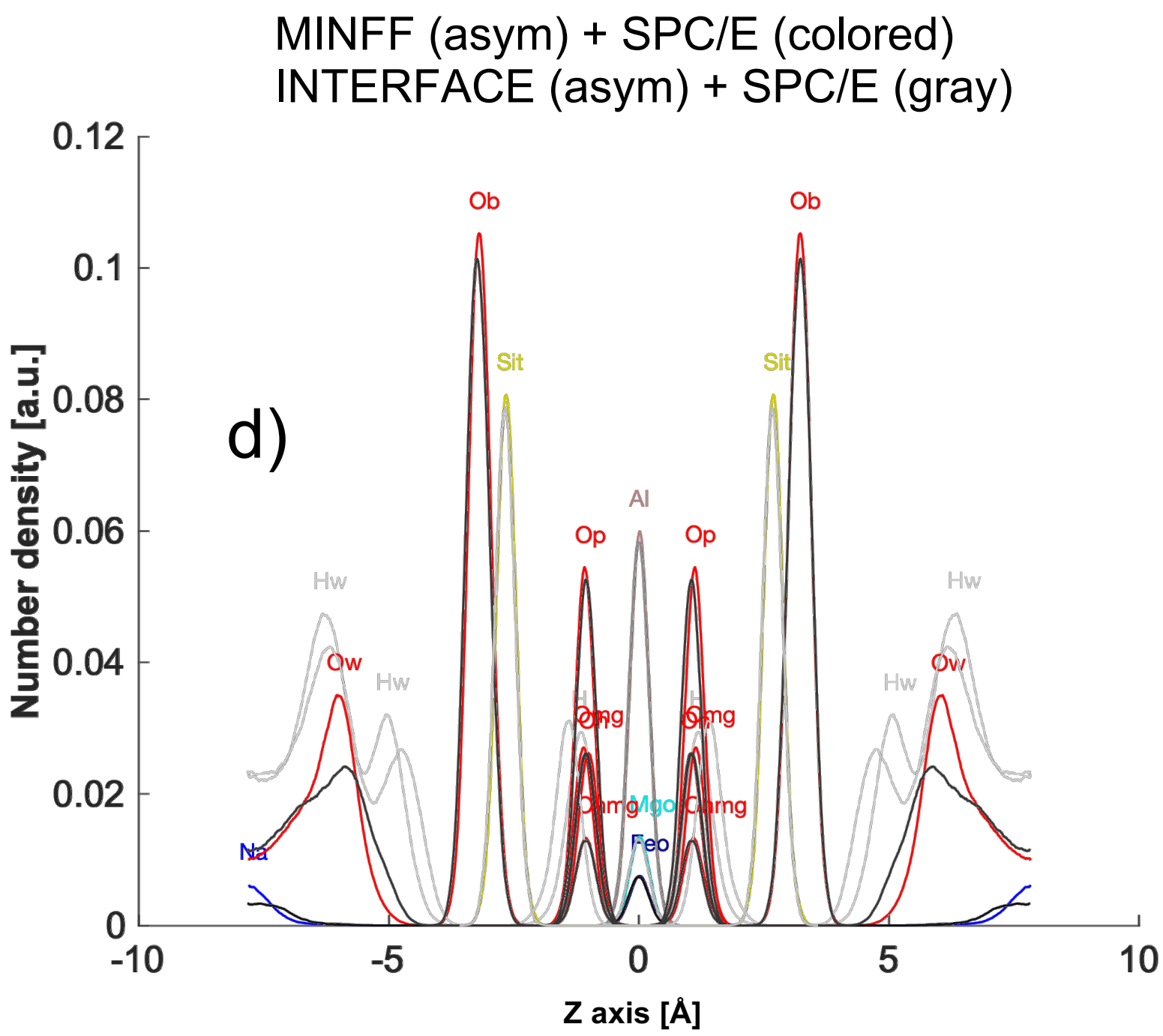
MINFF (asym) + SPC/E (colored)
INTERFACE (asym) + SPC/E (gray)
d)
Number density [a.u.]
Z axis [Å]
Ob
Sit
Op
Al
Hw
Ow
Na
Feo
Mgo

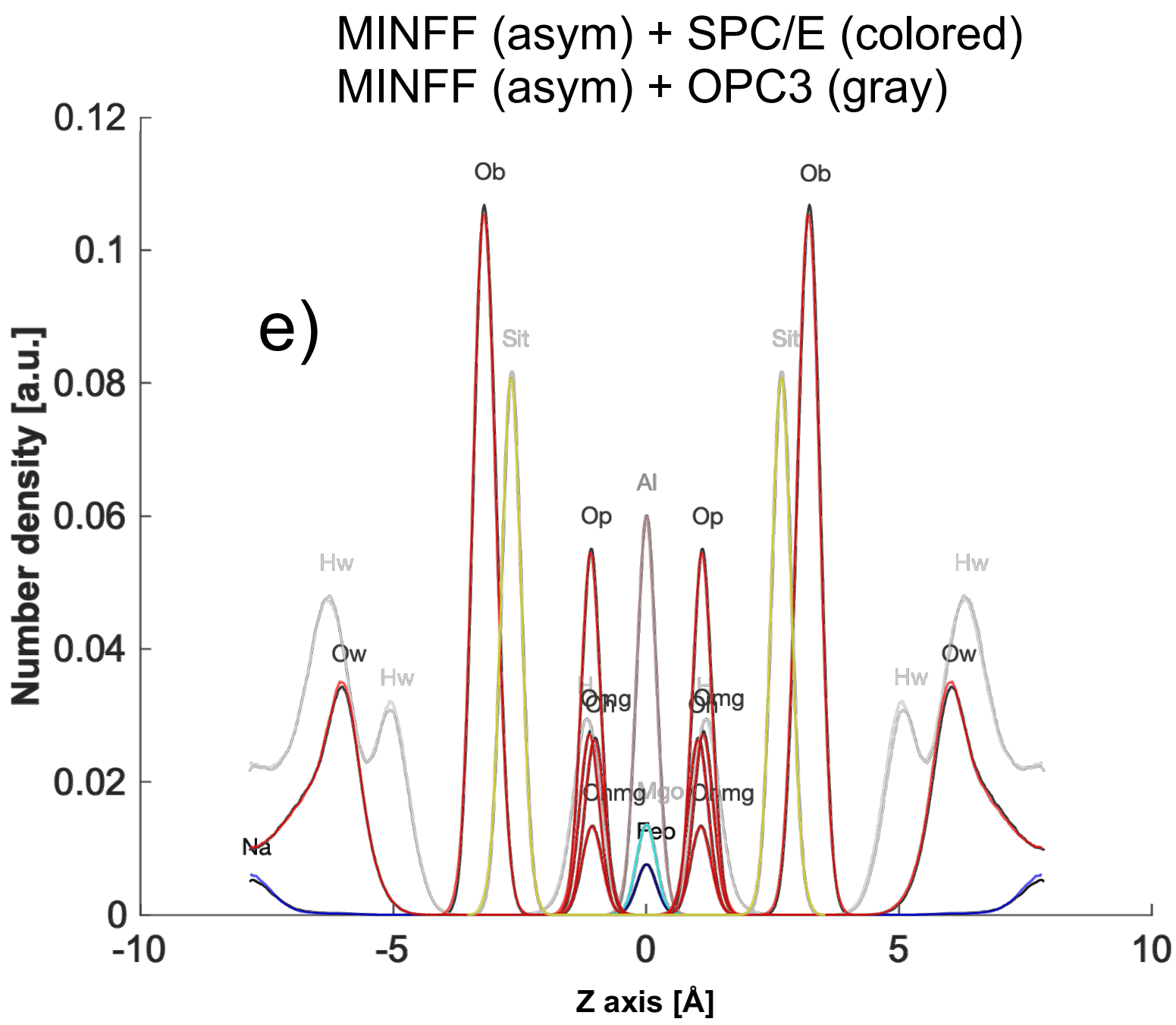

**Figure S6**. Comparison of overlapping density profiles between pairs of 15.60 Å 2W Na-Montmorillonite systems modelled with GMINFF, CLAYFF, and INTERFACE, having either hexagonal and symmetrical (sym) basal plane cavities or the experimental asymmetrical and ditrigonal cavities (asym). a) native CLAYFF (sym) versus CLAYFF (asym) in SPC/E water. b) native GMINFF (asym) versus GMINFF (sym) in SPC/E water. c) native GMINFF (asym) versus native CLAYFF (sym) in SPC/E water. d) native GMINFF (asym) versus INTERFACE (asym) in SPC/E water. e) native GMINFF (asym) in SPC/E and OPC3 water, respectively. Note that the density profiles are displayed either as coloured or with grey scaled curves as indicated in the figure titles.

## 4.4 Comparison of *00l* 1D-XRD profiles for the 2W hydration state in Na-Montmorillonite

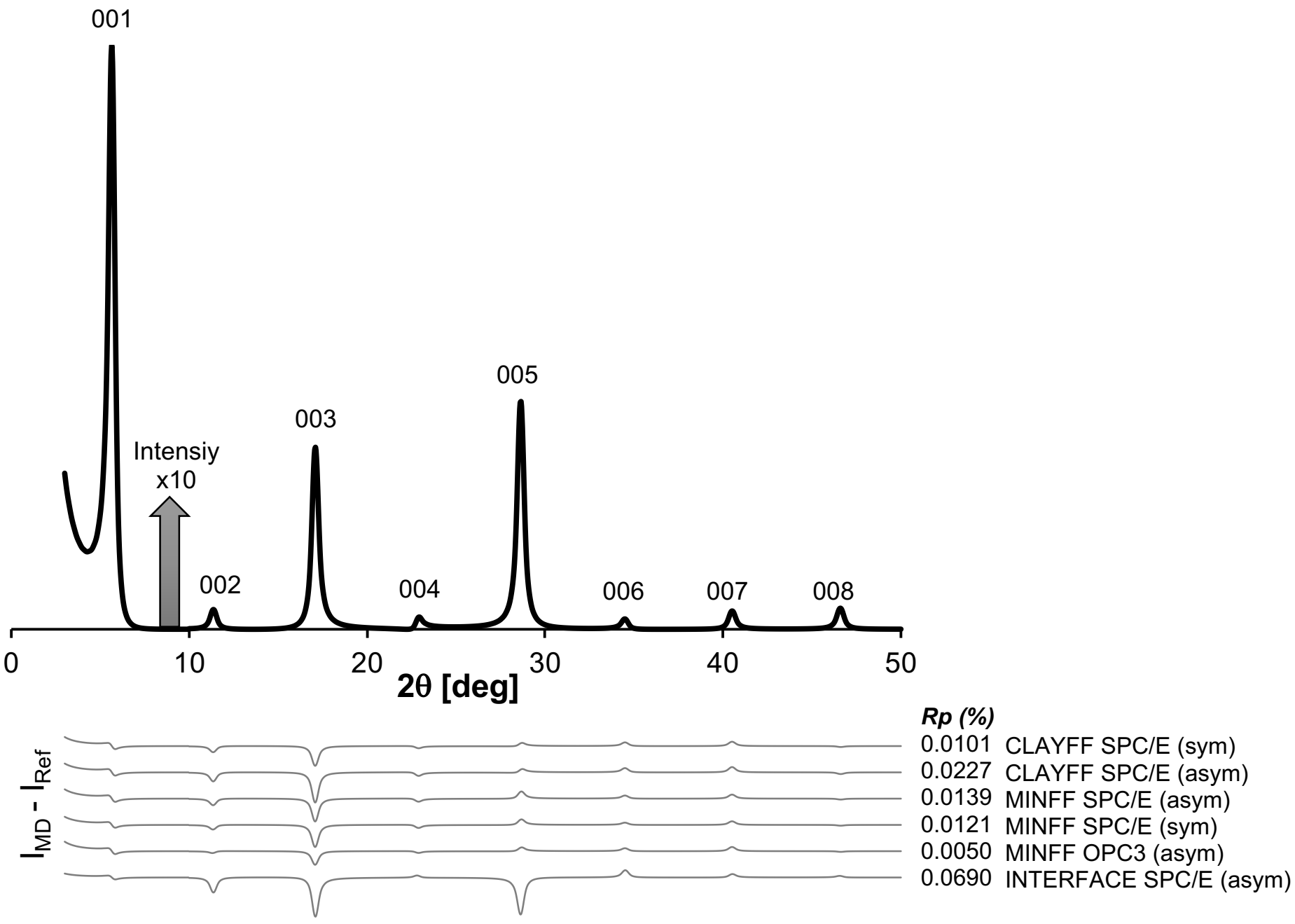

**Figure S7**. Comparison of theoretical 1D-XRD profiles (using 1.54187 Å) for different 15.60 Å 2W Na-Montmorillonite. The black profile shows the calculated reference random-mount powder XRD profile from a Reynolds smectite lattice and interlayer solute structure from MD simulations using CLAYFFmod and SPC/E water. Grey curves show corresponding differences to XRD profiles based on interlayer structures from MD simulations using (in order) i) CLAYFF w. SPC/E and native symmetrical hexagonal cavities, ii) CLAYFF w. SPC/E and ditrigonal asymmetrical cavities, iii) GMINFF w. SPC/E and native ditrigonal asymmetrical cavities, iv) GMINFF w. SPC/E and symmetrical hexagonal cavities, v) GMINFF w. OPC3 and native ditrigonal asymmetrical cavities, and vi) INTERFACE with SPC/E, based on the Rothbauer (1971) structure. ***Rp*** (in %) indicates the accuracy relative to the reference profile.

## 5. Umbrella sampling simulations with GMINFF vs. water models

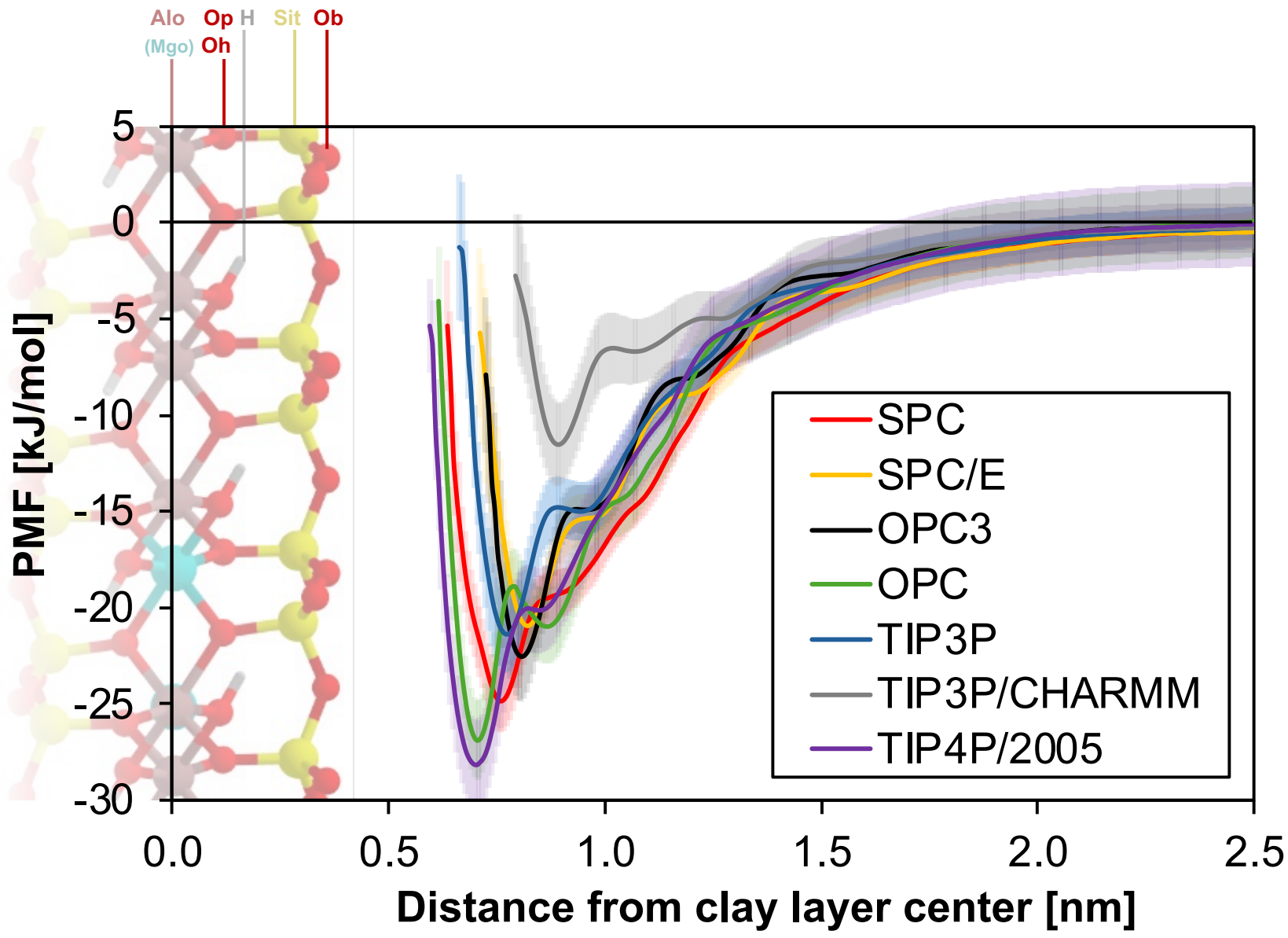

**Figure S8.** Sensitivity of the Ciprofloxacin adsorption free energy to the choice of water model, shown as potentials of mean force against distance from the centre of a Na-Montmorillonite layer. All profiles were computed with GMINFF and the native asymmetric ditrigonal cavities at $k_a$= 500 kJ·mol$^{-1}$·rad$^{-2}$, with zwitterionic Ciprofloxacin described by CGENFF, so that the water model is the only quantity varied. The seven models compared are SPC, SPC/E, OPC3, OPC, TIP3P, TIP3P/CHARMM, and TIP4P/2005, and the shaded bands indicate the uncertainty of each profile (n=10).

**Table S4**. Global PMF minimum of the Ciprofloxacin at the Montmorillonite surface modelled with GMINFF and different water models, its distance from the centre of the Na-Montmorillonite layer as shown in Figure S8, including the corresponding sorption coeff. $K_d$.

| | SPC | SPC/E | OPC3 | OPC | TIP3P | TIP3P/CHARMM | TIP4P/2005 |
|---|---|---|---|---|---|---|---|
| Min. PMF [kJ/mol] | -24.85 ±1.65 | -20.94 ±1.34 | -22.53 ±1.24 | -26.89 ±2.06 | -21.38 ±2.03 | -11.52 ±1.96 | -28.16 ±2.36 |
| Min. distance [nm] | 0.76 | 0.82 | 0.81 | 0.70 | 0.77 | 0.90 | 0.70 |
| $K_d$ [cm$^3$/g] | 1623.7 | 342.0 | 536.6 | 2640.7 | 370.1 | 10.0 | 5203 |

## 6. Variance in the Lennard-Jones parameters

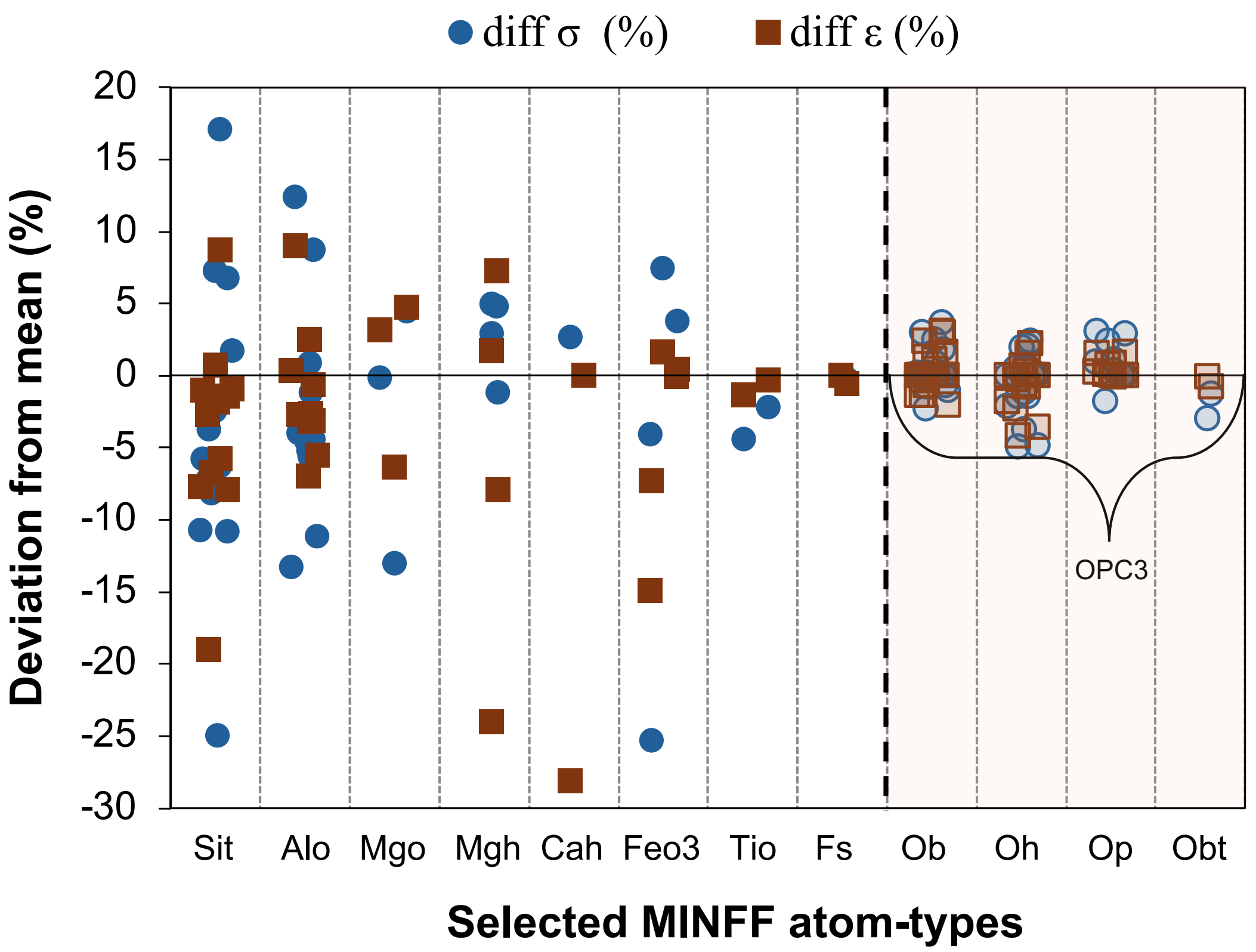

**Figure S9.** Spread of the tailored Lennard-Jones parameters across the training set at $k_a = 500$ kJ·mol$^{-1}$·rad$^{-2}$. Each point is one mineral and gives the deviation of the tailored $\sigma$ (blue circles) and $\varepsilon$ (red squares) from its reference value. For the metal atom types (filled circles, left of the dashed line) the reference is the GMINFF value and the tailored value is TMINFF-M. For the oxygen atom types (open squares, right of the dashed line) the reference is the OPC3 pair ($\sigma$ = 0.317427 nm, $\varepsilon$ = 0.68369 kJ/mol) used in GMINFF and TMINFF-M, and the tailored value is TMINFF-MO.

**Table S5.** Mean signed and root-mean-square (RMS) error of the nearest-neighbour M–O distance per atom-type and spread of the tailored Lennard-Jones parameters across the training set of minerals at $k_a$ = 500 kJ·mol$^{-1}$·rad$^{-2}$.[c]

| Atom-type | n[a] | Mean signed error (%) | | | RMS error (%) | | | Tailored CV[b] (%) | |
|---|---|---|---|---|---|---|---|---|---|
| | | GMINFF | TMINFF-M | TMINFF-MO | GMINFF | TMINFF-M | TMINFF-MO | $\sigma$ | $\varepsilon$ |
| Sit | 14 | −0.80 | −1.21 | −1.10 | 1.17 | 1.42 | 1.29 | 10.3 | 6.3 |
| Alo | 11 | 1.16 | 0.88 | 0.79 | 1.72 | 1.46 | 1.50 | 7.8 | 4.4 |
| Mgo | 3 | −0.39 | −0.60 | −0.59 | 1.00 | 1.10 | 1.10 | 9.3 | 6.0 |
| Mgh | 4 | −0.85 | −0.48 | −0.21 | 0.97 | 0.55 | 0.29 | 2.8 | 14.6 |
| Cah | 2 | −0.10 | −0.12 | 0.02 | 0.13 | 0.17 | 0.12 | 1.9 | 23.2 |
| Feo3 | 5 | −0.06 | −0.18 | −0.32 | 1.33 | 1.44 | 1.22 | 13.4 | 7.3 |
| Tio | 2 | 0.99 | 0.12 | 0.12 | 1.07 | 0.15 | 0.15 | 1.6 | 0.7 |
| Fs | 2 | −0.41 | −0.49 | −0.37 | 0.58 | 0.68 | 0.46 | 0.1 | 0.4 |
| Ob | 24 | −0.71 | −0.96 | −0.81 | 1.27 | 1.26 | 1.09 | 1.3 | 1.2 |
| Oh | 17 | 1.44 | 1.14 | 0.72 | 2.68 | 2.74 | 1.70 | 2.2 | 1.6 |
| Op | 12 | 0.26 | 0.03 | 0.42 | 1.15 | 0.90 | 1.14 | 1.4 | 0.6 |
| Obt | 2 | 0.96 | 0.97 | 0.27 | 0.97 | 1.06 | 0.36 | 1.2 | 0.5 |

[a] n is the number of training minerals that contain the atom-type. Distance errors are in % relative to the experimental reference structure, averaged over all minerals containing the atom-type.
[b] Coefficient of variation (CV = 100 × standard deviation / mean) of the tailored $\sigma$ and $\varepsilon$ across the minerals; TMINFF-M for the metal atom types and TMINFF-MO for the oxygen atom types, whose parameters are fixed at the OPC3 values in GMINFF and TMINFF-M. The tailored oxygen $\sigma$ stays within −5.0 to +3.7% of the OPC3 value (0.317427 nm) and $\varepsilon$ within −4.2 to +3.0% (OPC3: 0.68369 kJ·mol$^{-1}$).
[c] A constant offset per atom-type accounts for 36.8% (GMINFF), 33.4% (TMINFF-M), and 37.4% (TMINFF-MO) of the mean-square distance error at $k_a$ = 500 kJ·mol$^{-1}$·rad$^{-2}$. Across $k_a$ = 0, 250, 500, and 1500 kJ·mol$^{-1}$·rad$^{-2}$ this contribution range from 30.9 to 38.1% (GMINFF), 23.5 to 38.6% (TMINFF-M), and 30.5 to 42.0% (TMINFF-MO).